\newif\ifdraft
\newif\iffull
\newif\ifcomment
\newif\iflatexdiff
\newif\ifbibtex
\newif\ifpreprint
\latexdifftrue   
\preprinttrue    
\fulltrue        
\def\dvers{v3.4} 
\ifpreprint
\documentclass[ALICE,manyauthors,12pt]{cernphprep}
\usepackage[comma,square,numbers,sort&compress]{natbib}
\usepackage{hyperref}
\else
\documentclass[final,3p,12pt]{elsarticle}
\biboptions{comma,square,numbers,sort&compress}
\usepackage{hyperref}
\fi
\usepackage{graphicx}  
\usepackage{dcolumn}   
\usepackage{bm}        
\usepackage{amssymb}   
\usepackage{amsfonts}
\usepackage{graphics}
\usepackage{grffile} 
\usepackage{epsfig}
\usepackage[usenames]{color}
\iflatexdiff
\RequirePackage[normalem]{ulem}
\RequirePackage{color}\definecolor{RED}{rgb}{1,0,0}\definecolor{BLUE}{rgb}{0,0,1}

\fi
\ifpreprint
\else
 
\fi

\newcommand{\vn}           {\ensuremath{v_{n}}}
\newcommand{\vngf}         {\ensuremath{v_{n} \{ GF \} }}
\newcommand{\vnd}          {\ensuremath{V_{n\Delta}}}
\newcommand{\voned}          {\ensuremath{V_{1\Delta}}}
\newcommand{\ptt}          {\ensuremath{p_{T}^{t}}}
\newcommand{\pta}          {\ensuremath{p_{T}^{a}}}
\newcommand{\df}           {\ensuremath{\Delta\phi}}
\newcommand{\de}           {\ensuremath{\Delta\eta}}
\newcommand{\CF}           {\ensuremath{C \left(\Delta \phi \right)}}

\newcommand{\ZDC}          {\rm{ZDC}}
\newcommand{\ZDCs}         {\rm{ZDCs}}
\newcommand{\SPD}          {\rm{SPD}}

\newcommand{\TPC}          {\rm{TPC}}
\newcommand{\VZERO}        {\rm{VZERO}}
\newcommand{\VZEROA}       {\rm{VZERO-A}}
\newcommand{\VZEROC}       {\rm{VZERO-C}}

\newcommand{\PbPb}         {\mbox{Pb--Pb}}

\newcommand{\pt}           {\ensuremath{p_{T}}{ }}

\newcommand{\snn}          {\ensuremath{\sqrt{s_{\rm NN}}}}

\newcommand{\abs}[1]       {\ensuremath{\left|#1\right|}}

\newcommand{\dd}           {\ensuremath{\rm{d}}}

\newcommand{\Fig}[1]       {Fig.~\ref{#1}}
\newcommand{\Figure}[1]    {Figure~\ref{#1}}
\newcommand{\Eq}[1]        {Eq.~\ref{#1}}
\newcommand{\Equation}[1]  {Equation~\ref{#1}}
\newcommand{\Ref}[1]       {Ref.~\cite{#1}}

\newcommand{\warn}[1]      {{\small\textbf{(!\footnote{\textbf{(!)}~#1})}}\marginpar{\textbf{---}}}

\newcommand*{\doi}[1]{\href{http://dx.doi.org/#1}{doi: #1}}
\newcommand {\arxiv}[1]    {\href{http://www.arxiv.org/abs/#1}{\mbox{arXiv:#1}}}

\graphicspath{{./img/}}
\ifdraft
\usepackage{lineno}
\linenumbers
\modulolinenumbers[5]
\usepackage{fancyhdr}
\else
\renewcommand{\warn}[1]{}

\fi
\begin{document}
\newlength{\figlen}
\setlength{\figlen}{\linewidth}
\ifpreprint
\setlength{\figlen}{0.95\linewidth}
\begin{titlepage}
\PHnumber{2011-152}                   
\PHdate{11 Sep 2011}                  
\title{Harmonic decomposition of two-particle angular correlations\\
       in Pb--Pb collisions at $\mathbf{\sqrt{s_{\rm NN}} = 2.76}$ TeV}
\ShortTitle{Harmonic decomposition of two particle correlations}
\Collaboration{ALICE Collaboration%
         \thanks{See Appendix~\ref{app:collab} for the list of collaboration members}}
\ShortAuthor{ALICE Collaboration} 
\ifdraft
\begin{center}
\today\\ \color{red}DRAFT \dvers\ \hspace{0.3cm} \$Revision: 352 $\color{white}:$\$\color{black}\vspace{0.3cm}
\end{center}
\fi
\else
\begin{frontmatter}
\title{Harmonic decomposition of two particle angular correlations\\
       in Pb--Pb collisions at $\sqrt{s_{\rm NN}} = 2.76$ TeV}
\iffull
\input{authors-plb.tex}
\else
\ifdraft
\author{ALICE Collaboration \\ \vspace{0.3cm} 
\today\\ \color{red}DRAFT \dvers\ \hspace{0.3cm} \$Revision: 352 $\color{white}:$\$\color{black}}
\else
\author{ALICE Collaboration}
\fi
\fi
\fi
\begin{abstract}
  Angular correlations between unidentified charged trigger~($t$) and
  associated~($a$) particles are measured by the ALICE experiment in
  \PbPb\ collisions at $\snn=2.76$~TeV for transverse momenta $0.25 <
  p_{T}^{t,\, a} < 15$~GeV/$c$, where $p_{T}^t > p_{T}^a$.  The shapes
  of the pair correlation distributions are studied in a variety of
  collision centrality classes between 0 and 50\% of the total
  hadronic cross section for particles in the pseudorapidity interval
  $|\eta| < 1.0$. Distributions in relative azimuth $\Delta\phi \equiv
  \phi^t - \phi^a$ are analyzed for $|\Delta\eta| \equiv |\eta^t -
  \eta^a| > 0.8$, and are referred to as ``long-range correlations''.
  Fourier components $V_{n\Delta} \equiv \langle
  \cos(n\Delta\phi)\rangle$ are extracted from the long-range
  azimuthal correlation functions.%
  If particle pairs are correlated to one another through their
  individual correlation to a common symmetry plane, 
  then the pair anisotropy $\vnd(p_{T}^t, p_{T}^a)$ is fully described
  in terms of single-particle anisotropies $v_n (\pt)$ as
  $V_{n\Delta}(p_{T}^t, p_{T}^a) = v_n(p_{T}^t) \, v_n(p_{T}^a)$. This
  expectation is tested for $1 \leq n \leq 5$ by applying a global fit
  of all $\vnd (p_{T}^t, p_{T}^a)$ to obtain the best values $\vngf
  (\pt)$. It is found that for $2 \leq n \leq 5$, the fit agrees well
  with data up to $\pta \sim 3$-4 GeV/$c$, with a trend of increasing
  deviation as $p_{T}^t$ and $p_{T}^a$ are increased or as collisions
  become more peripheral. This suggests that no pair correlation
  harmonic can be described over the full $0.25 < p_{T} < 15$~GeV/$c$
  range using a single $\vn(\pt)$ curve; such a description is however
  approximately possible for $2 \leq n \leq 5$ when $\pta < 4$
  GeV/$c$.
  For the $n=1$ harmonic, however, a single $v_1(\pt)$ curve is not
  obtained even within the reduced range $ \pta < 4$ GeV/$c$.
  \ifdraft \ifpreprint
\end{abstract}
\end{titlepage}
\setcounter{page}{2}
\else
\end{abstract}
\end{frontmatter}
\newpage
\fi
\fi
\ifdraft
\thispagestyle{fancyplain}
\else
\end{abstract}
\ifpreprint
\end{titlepage}
\else
\end{frontmatter}
\setcounter{page}{2}
\fi
\fi
\setcounter{page}{2}

\iffull
\ifpreprint
\else
\clearpage
\fi
\fi

\section{Introduction} \label{sec:intro} Ultra-relativistic collisions
of large nuclei at the Large Hadron Collider~(LHC) and at the
Relativistic Heavy Ion Collider~(RHIC) enable the study of
strongly-interacting nuclear matter at extreme temperatures and energy
densities. One key piece of evidence for the formation of dense
partonic matter in these collisions is the observation of particle
momentum anisotropy in directions transverse to the
beam~\cite{Arsene:2004fa,Back:2004je,Adams:2005dq,Adcox:2004mh,Aamodt:2010pa,Collaboration:2011yk}.
One powerful technique to characterize the properties of the medium is
with two-particle correlations~\cite{Adler:2002tq, Adams:2005ph,
  Adare:2006nr, Adams:2006yt, Alver:2008gk, Adare:2008cqb,
  Abelev:2009qa, Alver:2009id, Adare:2010ry, Khachatryan:2010gv,
  ATLAS-CONF-2011-074, Chatrchyan:2011ek}, which measure the
distributions of angles $\df$ and/or $\de$ between particle pairs
consisting of a ``trigger'' at transverse momentum $\ptt$ and an
``associated'' partner at $\pta$.

In proton-proton collisions, the full ($\df$, $\de$) correlation
structure at ($\df$, $\de$) $\approx$ $(0, 0)$ is dominated by the
``near-side'' jet peak, where trigger and associated particles
originate from a fragmenting parton, and at $\df \approx \pi$ by the
recoil or ``away-side'' jet.  The away-side peak is broader in $\de$,
due to the longitudinal momentum distribution of partons in the
colliding nuclei. In central nucleus--nucleus collisions at RHIC, an
additional ``ridge'' feature is observed at
$\df\approx0$~\cite{Abelev:2009qa, Alver:2009id}, which has generated
considerable theoretical interest~\cite{Voloshin:2006ei,
  Armesto:2004pt, Chiu:2005ad, Romatschke:2006bb, Majumder:2006wi,
  Shuryak:2007fu, Wong:2008yh, Dumitru:2008wn, Gavin:2008ev,
  Dusling:2009ar, Hama:2009vu} since its initial observation.  With
increasing $\pt$, the contribution from the near-side jet peak
increases, while the ridge correlation maintains approximately the
same amplitude.  The recoil jet correlation is significantly weaker
than that of the near side, because of kinematic
considerations~\cite{Morsch:2006pf} and also because of partonic
energy loss. When both particles are at high transverse momenta ($\pta
\gtrsim 6$ GeV/$c$), the peak shapes appear similar to the
proton-proton case, albeit with a more suppressed away side. This
away-side correlation structure becomes broader and flatter than in
proton-proton collisions as the particle $\pt$ is decreased. In fact,
in very central events ($\approx 0$--$2\%$), the away side exhibits a
concave, doubly-peaked feature at $\abs{\df-\pi}\approx
60^\circ$~\cite{Aamodt:2011vk}, which also extends over a large range
in $|\de|$~\cite{ATLAS-CONF-2011-074, Chatrchyan:2011ek}.  The latter
feature has been observed previously at RHIC~\cite{Abelev:2009qa,
  Alver:2009id, Adare:2008cqb}, but only after subtraction of a
correlated component whose shape was exclusively attributed to
elliptic flow.

However, recent studies suggest that fluctuations in the initial state
geometry can generate higher-order flow components~\cite{Manly:2005zy,
  Alver:2006wh, Mishra:2007tw, Mishra:2008dm, Takahashi:2009na,
  Sorensen:2010sqm, Alver:2010gr, Teaney:2010vd, Luzum:2010sp}.  The
azimuthal momentum distribution of the emitted particles is commonly
expressed as
\begin{equation} \label{eq:vn}
\frac{\dd N}{\dd\phi} \propto 1 + \sum_{n=1}^{\infty} 2 \vn(\pt)\,\cos\left( n(\phi-\Psi_n)\right)
\end{equation}
where $\vn$ is the magnitude of the $n^{\rm th}$ order harmonic
term relative to the angle of the initial-state spatial plane of symmetry
$\Psi_n$.  First measurements, in particular of $v_3$ and $v_5$
have been reported recently~\cite{Aamodt:2011vk, Adare:2011tg,
  ATLAS-CONF-2011-074}.

These higher-order harmonics contribute to the
previously-described structures observed in trigger-associated
particle correlations via the expression
\begin{equation} \label{eq:vnvn}
\frac{\dd N^{\rm pairs}}{\dd\df} \propto 1 + \sum_{n=1}^{\infty}
2 \vn(\ptt) \vn(\pta)\,\cos\left(n \df \right). 
\end{equation}

Similarly, the {\it measured} anisotropy from two-particle
correlations at harmonic order $n$ is given by $\vnd$:
\begin{equation} \label{eq:Vn}
\frac{\dd N^{\rm pairs}}{\dd\df} \propto 1 + \sum_{n=1}^{\infty}
2 \vnd(\ptt,\pta)\,\cos\left(n \df \right).
\end{equation}

In this article, we present a measurement of the $\vnd$ coefficients
from triggered, pseudo\-rapidity-separated ($|\de|>0.8$) pair
azimuthal correlations in \PbPb\ collisions in different centrality
classes and in several transverse momentum intervals. 
Details of the experimental setup and analysis are described in
sections~\ref{sec:data} and~\ref{sec:ana}, respectively.  The goal of
the analysis is to quantitatively study the connection between the
measured two-particle anisotropy $\vnd$ of \Eq{eq:Vn} and the
inclusive-particle harmonics of \Eq{eq:vnvn}. Specifically, we check
whether a set of single-valued $v_n(\pt)$ points can be identified
that describe the measured long-range anisotropy via the relation
$\vn(\ptt) \vn(\pta) = \vnd(\ptt,\pta)$. If so, $\vnd$ is said to {\it
  factorize} into single-particle Fourier coefficients within the
relevant $\ptt, \pta$ region. This relationship is tested for
different harmonics $n$ and in different centrality classes by
performing a global fit~($GF$) over all $p_{T}^{t,\, a}$ bins~(see
section~\ref{sec:globalfit}). The global fit procedure results in the
coefficients $\vngf(\pt)$ that best describe the anisotropy given by
the $\vnd(\ptt, \pta)$ harmonics as $\vngf(\ptt) \times
\vngf(\pta)$. The resulting $\vngf$ values for $1<n\leq 5$ are
presented in section~\ref{sec:vngf}.  
  A summary is given in section~\ref{sec:summary}.

\section{Experimental setup and data analysis} \label{sec:data} The
data used in this analysis were collected with the ALICE detector in
the first \PbPb\ run at the LHC (November 2010).  Charged particles
are tracked using the Time Projection Chamber~(\TPC), whose acceptance
enables particle reconstruction within $-1.0<\eta<1.0$. Primary vertex
information is provided by both the \TPC\ and the silicon pixel
detector (\SPD), which consists of two cylindrical layers of hybrid
silicon pixel assemblies covering $|\eta|<2.0$ and $|\eta|<1.4$ for
the inner and outer layers, respectively.  Two \VZERO\ counters, each
containing two arrays of 32 scintillator tiles and covering
$2.8<\eta<5.1$~(\VZEROA) and $-3.7<\eta<-1.7$~(\VZEROC), provide
amplitude and time information for triggering and centrality
determination. The trigger was configured for high efficiency to
accept inelastic hadronic collisions. The trigger is defined by a
coincidence of the following three conditions: i) two pixel hits in
the outer layer of the \SPD, ii) a hit in \VZEROA, and iii) a hit in
\VZEROC.

Electromagnetically induced interactions are rejected by requiring an
energy deposition above $500$~GeV in each of the Zero Degree
Calorimeters~(\ZDCs) positioned at $\pm$~114~m from the interaction
point.  Beam background events are removed using the \VZERO\ and \ZDC\
timing information.  The combined trigger and selection efficiency is
estimated from a variety of Monte Carlo~(MC) studies. This efficiency
ranges from 97\% to 99\% and has a purity of 100\% in the 0-90\%
centrality range.  The dataset for this analysis includes
approximately 13 million events.  Centrality was determined
by the procedure described in \Ref{Aamodt:2010cz}.  The centrality
resolution, obtained by correlating the centrality estimates of the
\VZERO, \SPD\ and \TPC\ detectors, is found to be about 0.5\% RMS for
the $0$--$10$\% most central collisions, allowing centrality binning
in widths of $1$ or $2$ percentiles in this range.

This analysis uses charged particle tracks from the ALICE \TPC\ having
transverse momenta from 0.25 to 15 GeV/$c$. The momentum resolution
$\sigma(\pt) / \pt$ rises with $\pt$ and ranges from $1$--$2$\% below
2 GeV/$c$ up to $10$--$15$\% near 15 GeV/$c$, with a negligible
dependence on occupancy. Collision vertices are determined using both
the \TPC\ and \SPD. Collisions at a longitudinal position greater than
10 cm from the nominal interaction point are rejected. The
closest-approach distance between each track and the primary vertex is
required to be within 3.2 (2.4) cm in the longitudinal (radial)
direction. At least 70 \TPC\ pad rows must be traversed by each track,
out of which 50 \TPC\ clusters must be assigned. In addition, a track
fit is applied requiring $\chi^2$ per TPC cluster $\leq$ 4 (with 2
degrees of freedom per cluster).

\begin{figure}[hbt] \centering 
  \includegraphics[width=0.45\textwidth]{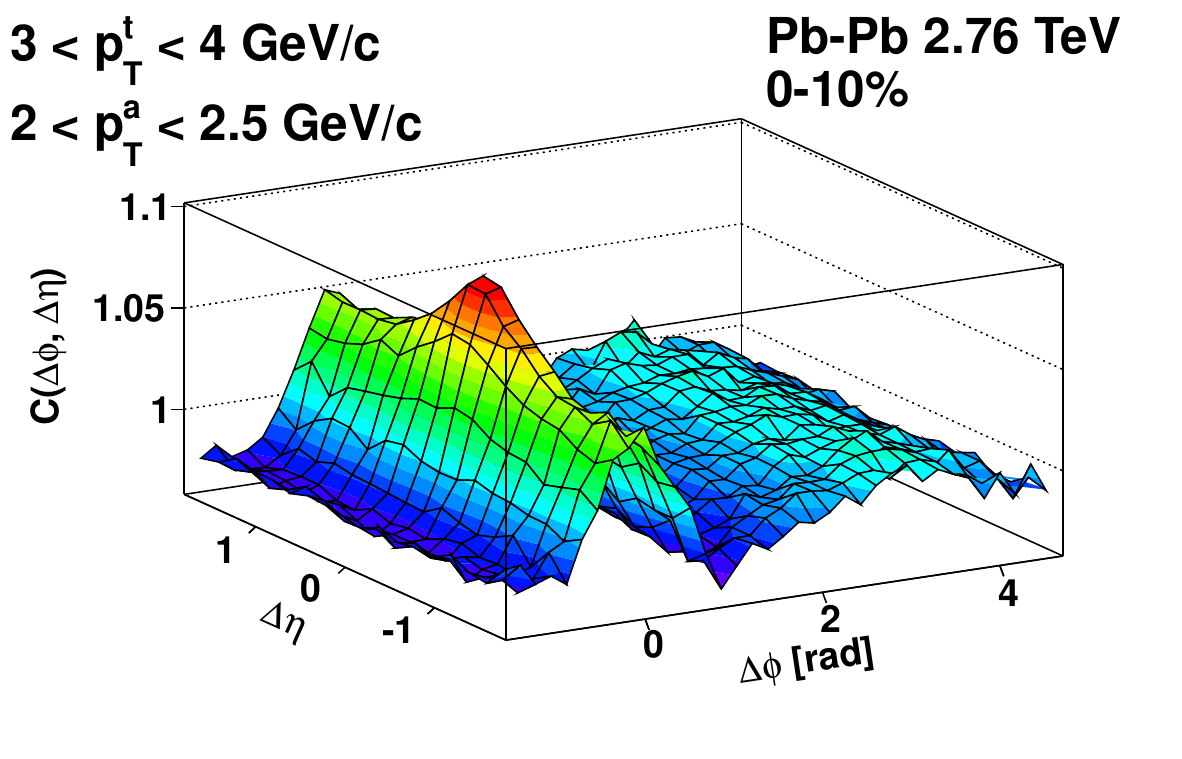}
  \qquad
  \includegraphics[width=0.45\textwidth]{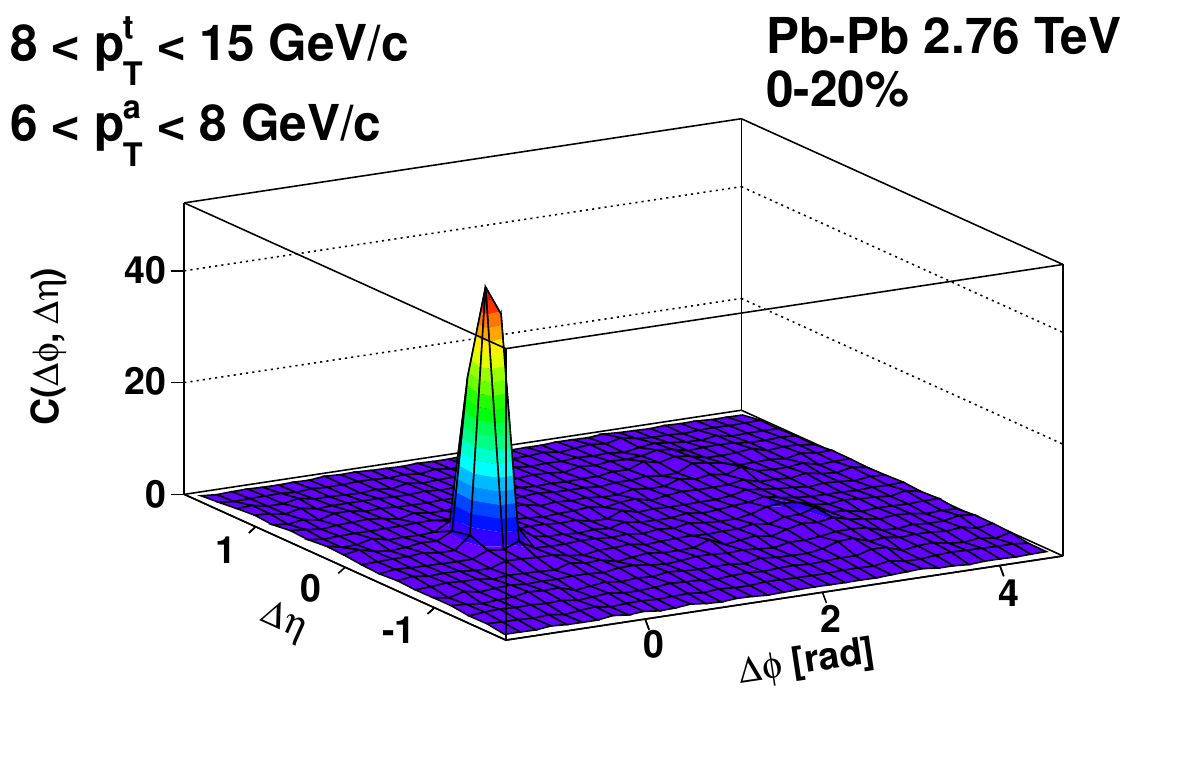}
  \caption[eta phi corr fns]{\label{fig:etaphi} Examples of
    two-particle correlation functions $C(\df, \de)$ for central
    \PbPb\ collisions at low to intermediate transverse momentum (left) 
    and at higher $\pt$ (right). Note the large difference in vertical
    scale between panels.}
\end{figure}

\section{Two-particle correlation function and Fourier
  analysis} \label{sec:ana} The two-particle correlation observable
measured here is the correlation function $C$(\df, \de), where the
pair angles $\df$ and $\de$ are measured with respect to the trigger
particle. The correlations induced by imperfections in detector
acceptance and efficiency are removed via division by a mixed-event
pair distribution $N_{\mathrm{mixed}} (\df, \de)$, in which a trigger
particle from a particular event is paired with associated particles
from separate events.  This acceptance correction procedure removes
structure in the angular distribution that arises from non-uniform
acceptance and efficiency, so that only physical correlations remain.
Within a given $\ptt$, $\pta$, and centrality interval,
the correlation function is defined as
\begin{equation} \label{eq:cfdef}
  C(\df, \de) \equiv \frac{N_{\mathrm{mixed}} } {N_{\mathrm{same}} }
    \times \frac{N_{\mathrm{same}} (\df, \de) } {N_{\mathrm{mixed}} (\df, \de) }.
\end{equation}
The ratio of mixed-event to same-event pair counts is included as a
normalization prefactor such that a completely uncorrelated pair
sample lies at unity for all angles. For $N_{\mathrm{mixed}} (\df,
\de)$, events are combined within similar categories of collision
vertex position so that the acceptance shape is closely reproduced,
and within similar centrality classes to minimize effects of residual
multiplicity correlations. To optimize mixing accuracy on the one hand
and statistical precision on the other, the event mixing bins vary
in width from $1$ to $10$\% in centrality and $2$ to $4$~cm in
longitudinal vertex position.
\begin{figure}[tbh] \centering 
  \includegraphics[width=0.36\textwidth]{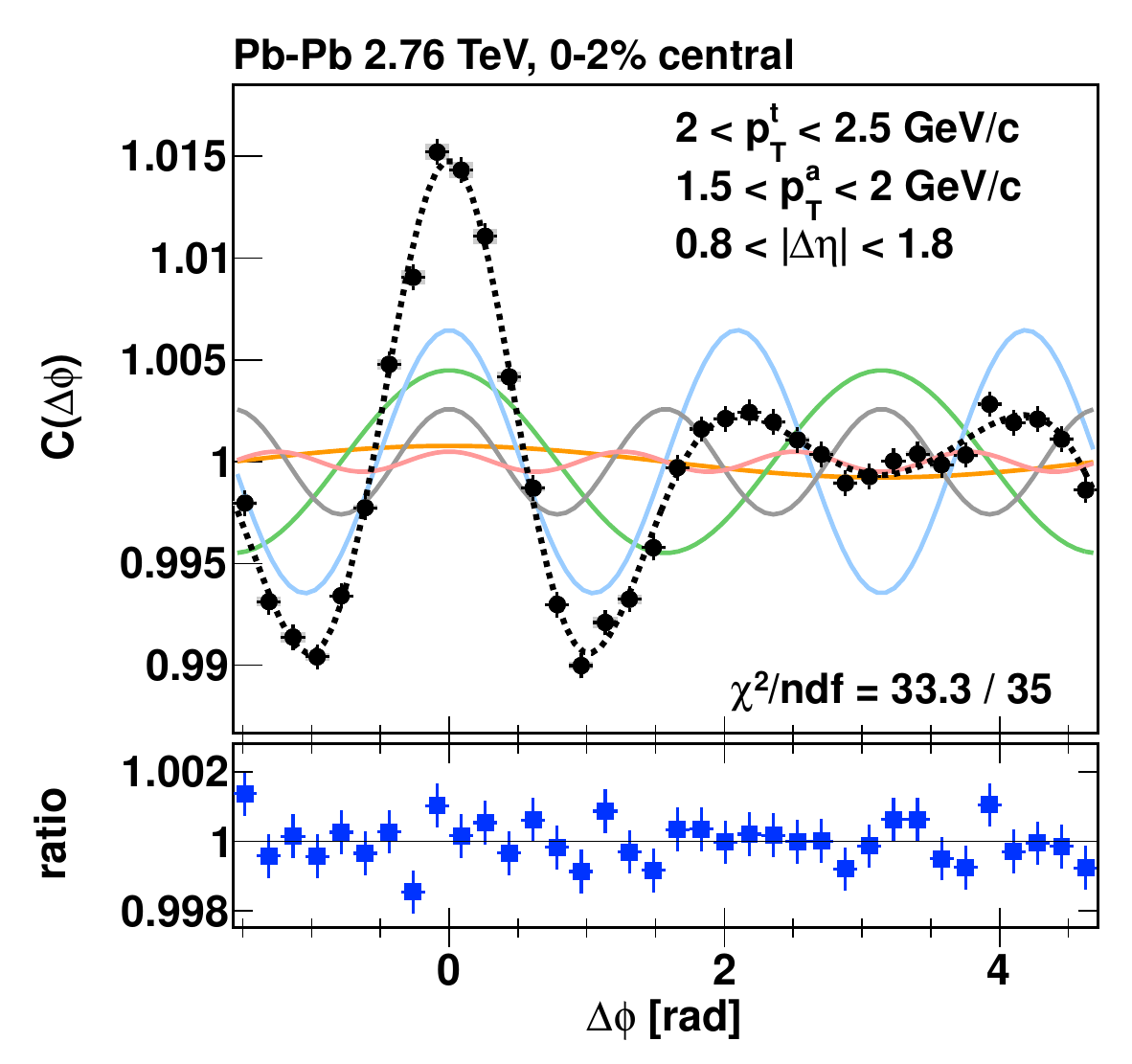}
  \includegraphics[width=0.31\textwidth]{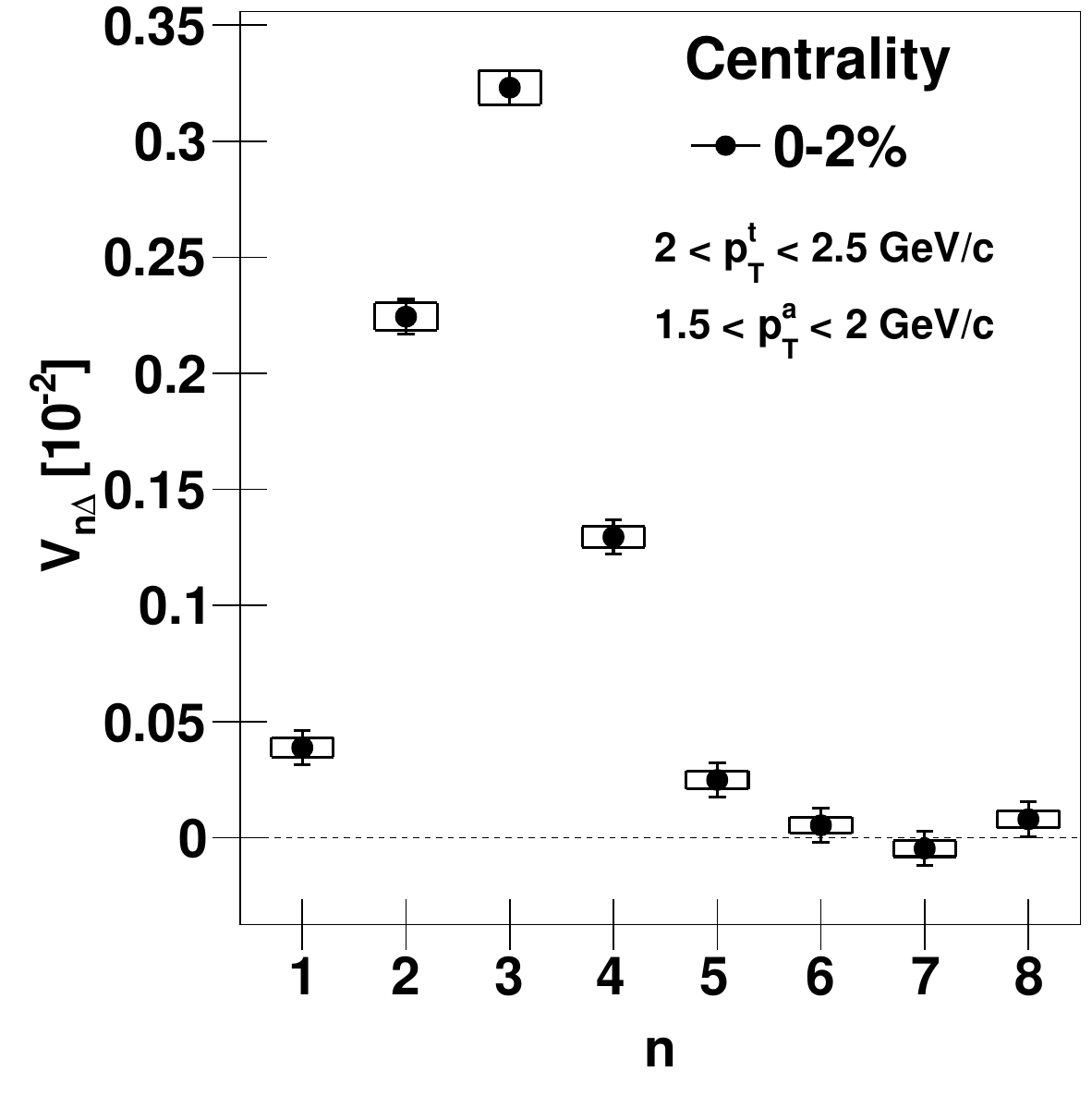}
  \includegraphics[width=0.31\textwidth]{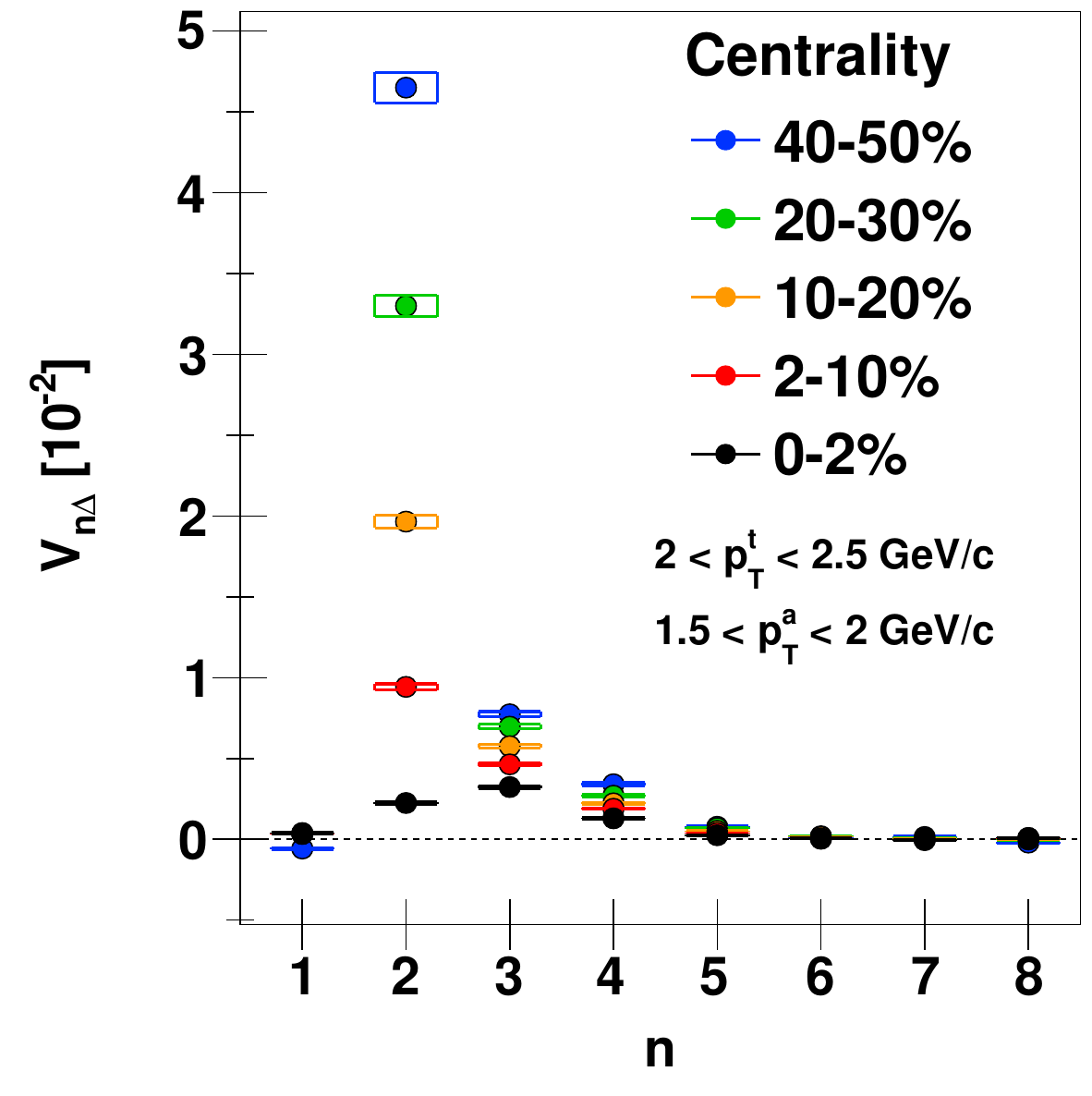}
  \caption[Corr fns]{\label{fig:cf_low} (Color online) Left: $\CF$ for
    particle pairs at $|\de| > 0.8$. The Fourier harmonics for
    $V_{1\Delta}$ to $V_{5\Delta}$ are superimposed in color. Their
    sum is shown as the dashed curve. The ratio of data to the $n \leq 5$
    sum is shown in the lower panel. Center: Amplitude of $\vnd$
    harmonics vs.\ $n$ for the same $\ptt$, $\pta$, and centrality
    class. Right: $\vnd$ spectra for a variety of centrality
    classes. Systematic uncertainties are represented with boxes (see
    section~\ref{sec:globalfit}), and statistical uncertainties are shown as
    error bars.}
\end{figure}

\begin{figure}[hbt] \centering 
  \includegraphics[width=0.36\textwidth]{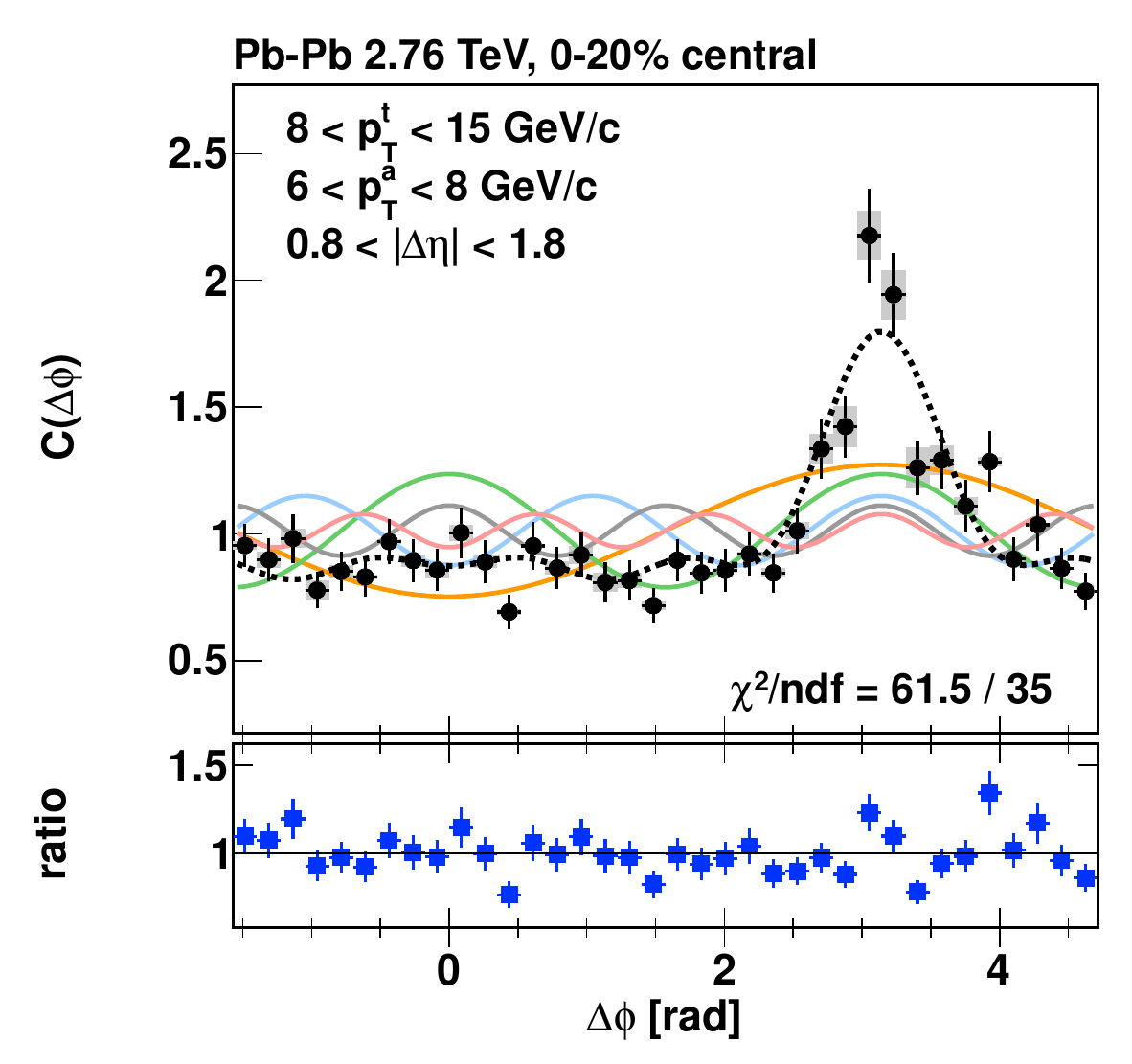}
  \quad
  \includegraphics[width=0.31\textwidth]{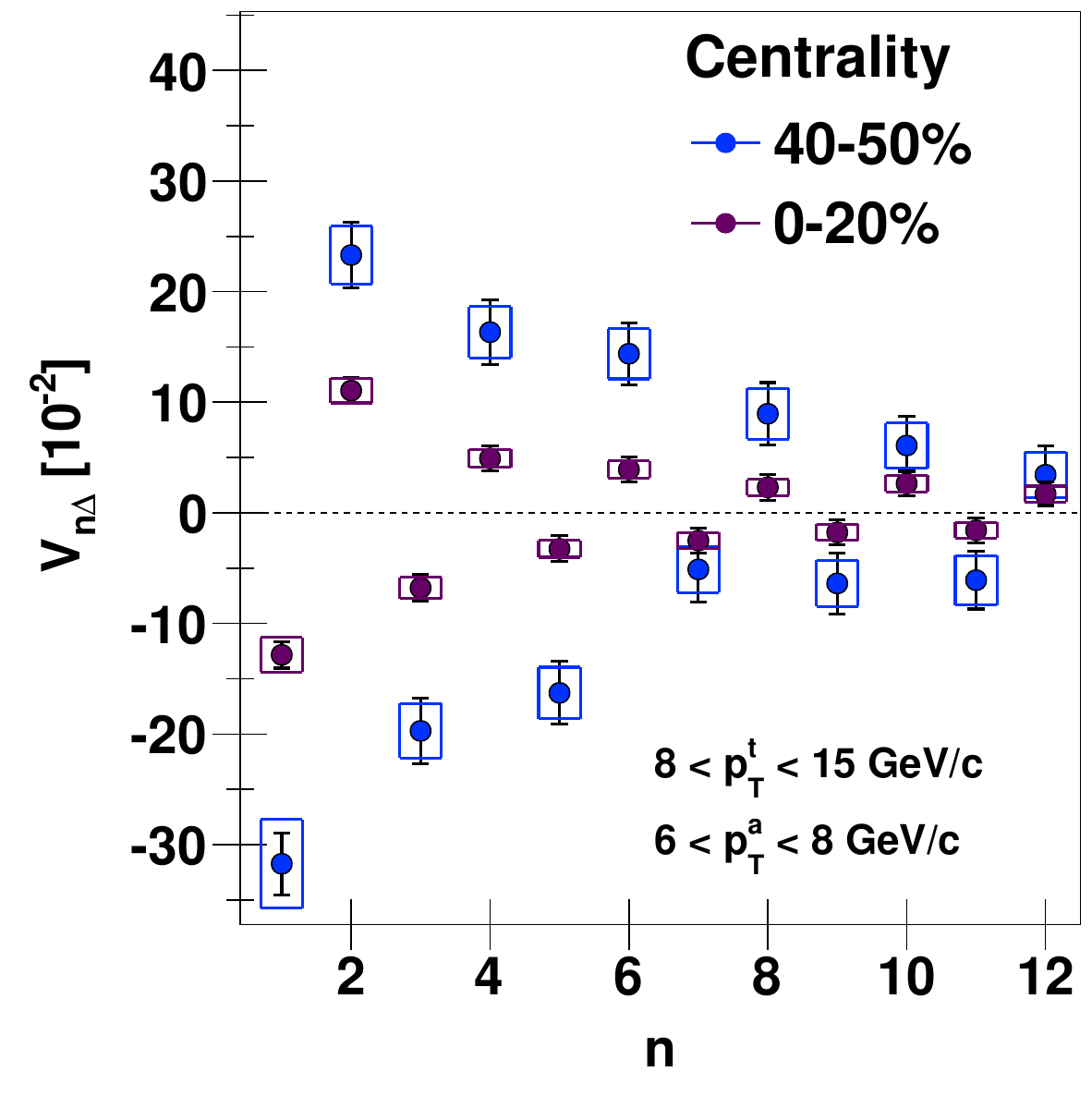}
  \caption[Power spectra]{\label{fig:cf_high} (Color online) Left:
    $\CF$ at $|\de| > 0.8$ for higher-$\pt$ particles than in
    \Fig{fig:cf_low}.  The Fourier harmonics $\vnd$ for $n \leq 5$ are
    superimposed in color.  Their sum is shown as the dashed
    curve. The ratio of data to the $n \leq 5$ sum is shown in the
    lower panel. Right: Amplitude of $\vnd$ harmonics vs.\ $n$ at the
    same $\ptt$, $\pta$ for two centrality bins. Systematic
    uncertainties are represented with boxes (see
    section~\ref{sec:globalfit}), and statistical uncertainties are
    shown as error bars.}
\end{figure}

It is instructive to consider the two examples of $C(\df, \de)$ from
\Fig{fig:etaphi} to be representative of distinct kinematic
categories. The first is the ``bulk-dominated'' regime, where
hydrodynamic modeling has been demonstrated to give a good description
of the data from heavy-ion collisions~\cite{Arsene:2004fa,
  Back:2004je, Adams:2005dq, Adcox:2004mh, Aamodt:2010pa}. We
designate particles with $\ptt$ (thus also $\pta$) below 3--$4$
GeV/$c$ as belonging to this region for clarity of discussion (see
\Fig{fig:etaphi}, left). A second category is the ``jet-dominated''
regime, where both particles are at high momenta ($\pta > 6$ GeV/$c$),
and pairs from the same di-jet dominate the correlation structures
(see \Fig{fig:etaphi}, right).

A major goal of this analysis is to quantitatively study the evolution
of the correlation shapes between these two regimes as a function of
centrality and transverse momentum. In order to reduce contributions
from the near-side peak, we focus on the correlation features at long
range in relative pseudorapidity by requiring $|\de| > 0.8$. This gap
is selected to be as large as possible while still allowing good
statistical precision within the TPC acceptance. The projection of
$C(\df, |\de| > 0.8)$ into $\df$ is denoted as $\CF$.

An example of $\CF$ from central \PbPb\ collisions in the
bulk-dominated regime is shown in \Fig{fig:cf_low} (left). The
prominent near-side peak is an azimuthal projection of the ridge seen
in \Fig{fig:etaphi}. In this very central collision class
($0$--$2$\%), a distinct doubly-peaked structure is visible on the
away side, which becomes a progressively narrower single peak in less
central collisions. We emphasize that no subtraction was performed on
$\CF$, unlike other jet correlation analyses~\cite{Adler:2002tq,
  Adams:2005ph, Adare:2006nr, Adams:2006yt, Alver:2008gk,
  Adare:2008cqb, Abelev:2009qa, Alver:2009id}.

A comparison between the left panels of~\Fig{fig:cf_low}
and~\Fig{fig:cf_high} demonstrates the change in shape as the
transverse momentum is increased. A single recoil jet peak at $\df
\simeq \pi$ appears whose amplitude is no longer a few percent, but
now a factor of 2 above unity. No significant near-side ridge is
distinguishable at this scale. The recoil jet peak persists even with
the introduction of a gap in $|\de|$ due to the distribution of
longitudinal parton momenta in the colliding nuclei.

The features of these correlations can be parametrized at
various momenta and centralities by decomposition into discrete
Fourier harmonics, as done~(for example) in~\cite{Alver:2010gr,
  Luzum:2010sp}.  Following the convention of those references, we
denote the two-particle Fourier coefficients as $\vnd$~(see
\Eq{eq:Vn}), which we calculate directly from $\CF$ as
\begin{equation} \label{eq:vnd}
\vnd \equiv
\langle \cos \left(n\Delta\phi\right) \rangle 
= {\displaystyle \sum_i^{} C_i \cos (n\df_i) } \left/ {\displaystyle
    \sum _i^{} C_i \,.} \right.
\end{equation}
Here, $C_i$ indicates that the $\CF$ is evaluated at $\df_i$.  Thus
$\vnd$ is independent of the normalization of \CF. The
$\vnd$ harmonics are superimposed on the left panels of
\Fig{fig:cf_low} and \Fig{fig:cf_high}. In the right panels, the
$\vnd$ spectrum is shown for the same centrality and momenta, with
additional centrality classes included to illustrate the centrality
dependence. The systematic uncertainties in these figures are
explained in section~\ref{sec:globalfit}.

\begin{figure}[thb!] \centering 
  \includegraphics[width=0.99\textwidth]{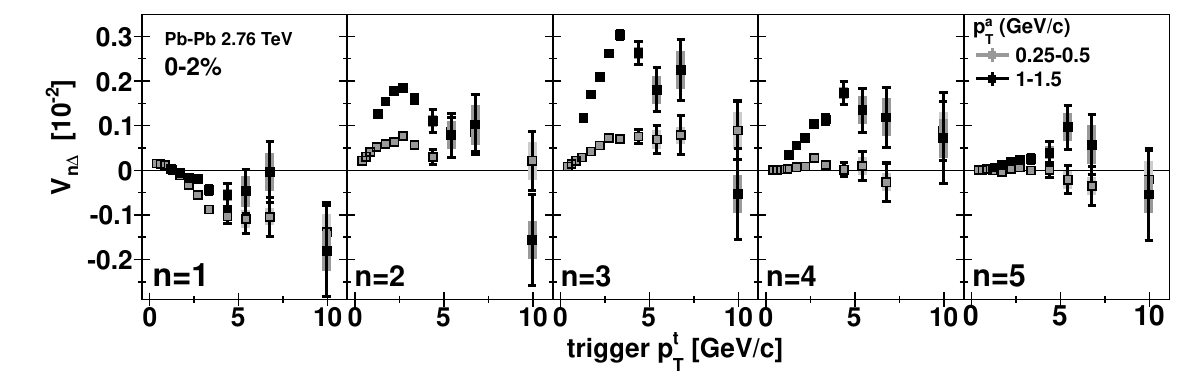}
  \includegraphics[width=0.99\textwidth]{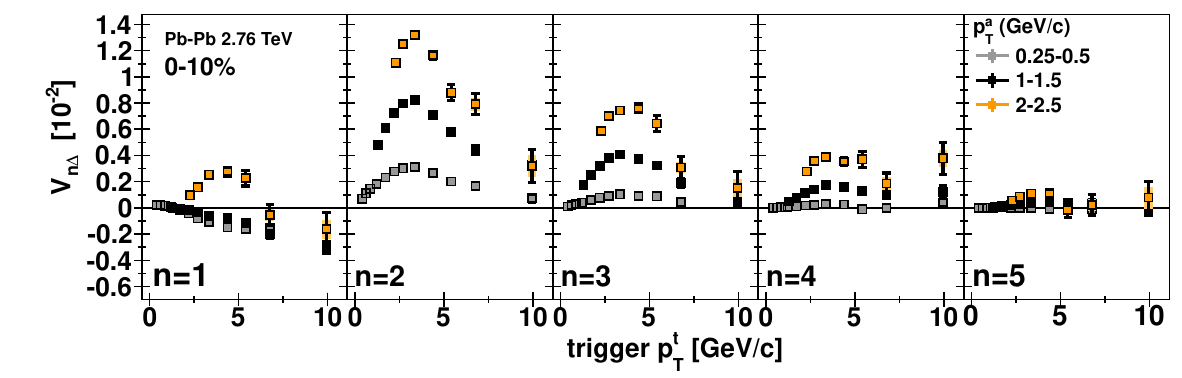}
  \includegraphics[width=0.99\textwidth]{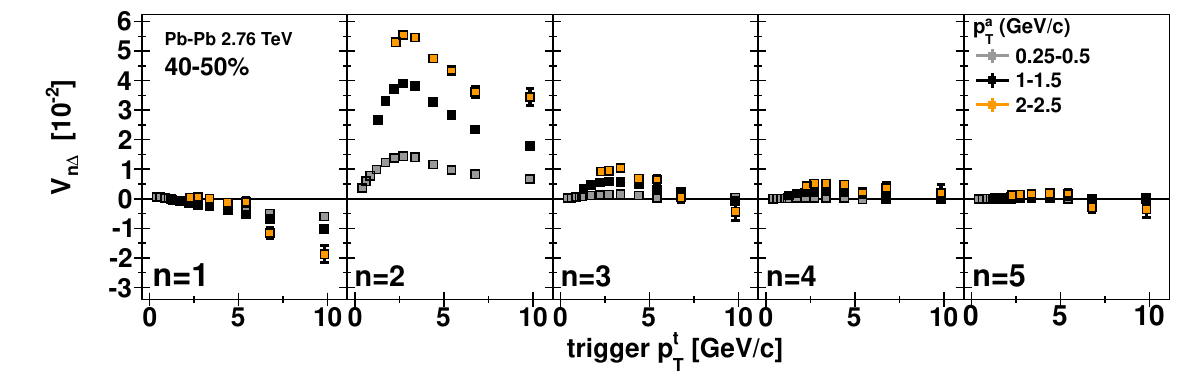}
  \caption[V2DeltaHi]{\label{fig:VnDelta} $V_{n\Delta}$ coefficients
    as a function of $\ptt$ for the $0$--$2$\%, $0$--$10$\%, and $40$--$50$\% most
    central \PbPb\ collisions (top to bottom).}
\end{figure}


In the bulk-dominated momentum regime and for central
collisions~(\Fig{fig:cf_low}), the first few Fourier harmonics are
comparable in amplitude, with the notable exception of
$V_{1\Delta}$. The first 5 combined harmonics reproduce $\CF$ with
high accuracy, as shown in the ratio between the points and the
component sum. For less central collisions, $V_{2\Delta}$ increasingly
dominates.  In the high-$\pt$ regime~(\Fig{fig:cf_high}), the jet peak
at $\df = \pi$ is the only prominent feature of the correlation
function. The even (odd) harmonics take positive (negative) values
which diminish in magnitude with increasing $n$, forming a pattern
distinct from the low-$\pt$ case. The dependence of the values on $n$
in the left panel of~\Fig{fig:cf_high} is approximately consistent
with a Gaussian function centered at $n=0$, as expected for the
Fourier transform of a Gaussian distribution of width $\sigma_{n} =
1/\sigma_{\df}$ centered at $\df = \pi$.  In this case, the sum of the
first 5 harmonics does not reproduce $\CF$ with the accuracy of the
low-\pt case, as suggested by the larger $\chi^2$ value (61.5/35
compared to 33.3/35). Although not shown, it was found that including
additional harmonics significantly improves the $\chi^2/NDF$ measure
for high-$\pta$ correlations, but adding higher orders in the
bulk-dominated case has only a modest effect. For example, if curves
composed of the lowest ten harmonics are used, the $\chi^2$ value
drops by only about 10\% to 30.0/35 in~\Fig{fig:cf_low}, but by over
25\% to 45.7/35 in~\Fig{fig:cf_high}. We note that $v_2$ is not the
dominant coefficient in~\Fig{fig:cf_high}; instead, its magnitude fits
into a pattern without significant dependence on collision geometry,
as suggested by the continuous decrease with increasing $n$ for both
$0$--$20$\% and $40$--$50$\% central events. This suggests that the
$n$ spectrum is driven predominantly by intra-jet correlations on the
recoil side, as expected from proton-proton correlations at similar
particle momenta.

\Figure{fig:VnDelta} shows the $\vnd$ coefficients as a function of
trigger \pt for a selection of associated \pt values. For $n \geq 2$,
$\vnd$ reaches a maximum value at $\ptt \simeq 3$--$4$ GeV/$c$,
decreasing toward zero (or even below zero for odd $n$) as $\ptt$
increases. This rapid drop of the odd coefficients at
high $\ptt$ provides a complementary picture to the $n$ dependence of
$\vnd$ shown in~\Fig{fig:cf_high}.

\section{Factorization and the global fit} \label{sec:globalfit} The
trends in $\ptt$ and centrality in \Fig{fig:VnDelta} are reminiscent
of previous measurements of $\vn$ from anisotropic flow
analyses~\cite{Aamodt:2011vk, Adare:2011tg, ATLAS-CONF-2011-074}.
This is expected if the azimuthal anisotropy of final state particles
at large $|\de|$ is induced by a collective response to initial-state
coordinate-space anisotropy from collision geometry and
fluctuations~\cite{Sorensen:2010sqm}. In
such a case, $\CF$ reflects a mechanism that affects all particles in
the event, and $\vnd$ depends only on the single-particle azimuthal
distribution with respect to the $n$-th order symmetry plane
$\Psi_n$. Under these circumstances $V_{n\Delta}$ factorizes as
\begin{eqnarray}\label{eq:fac}
  V_{n\Delta} (\ptt, \pta) &=& \langle\langle e^{in(\phi_a -
    \phi_t)} \rangle \rangle
  \nonumber \\ 
  &=& \langle \langle e^{in(\phi_a - \Psi_n)} \rangle \langle e^{-in(\phi_t -
    \Psi_n)} \rangle \rangle \nonumber \\
  &=& \langle v_n\{2\}(\ptt) \, v_n\{2\}(\pta) \rangle.
\end{eqnarray}

Here, $\langle \, \rangle$ indicates an averaging over events,
$\langle\langle \, \rangle\rangle$ denotes averaging over both
particles and events, and $\vn\{ 2\}$ specifies the use of a
two-particle measurement to obtain $\vn$. 

Equation~\ref{eq:fac} represents the factorization of $\vnd (\ptt,
\pta)$ into the event-averaged product $\langle v_n\{2\}(\ptt)$ $
v_n\{2\}(\pta) \rangle$, which includes event-by-event
fluctuations. Consistency with~\Eq{eq:fac} suggests that a large
fraction of the particle pairs are correlated through their individual
correlation with a common plane of symmetry. For example, symmetry
planes for particle pairs at $\ptt$ and $\pta$ may develop at a
harmonic order $n$ from collision geometry, initial state density
fluctuations, or from an axis formed by (di-)jet fragmentation. If a
single-valued $v_n(\pt)$ curve on an interval containing $\ptt$ and
$\pta$ can reproduce the magnitude of any $\vnd(\ptt, \pta)$, then
$\vnd$ factorizes within the $(\ptt, \pta)$ region.

\begin{figure}[tbh] \centering 
  \includegraphics[width=0.98\textwidth]{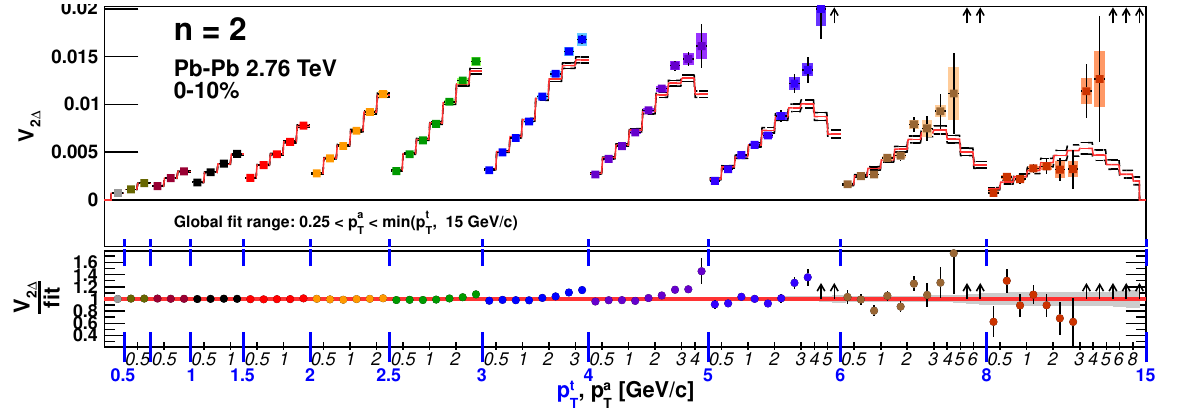} \\
  \vspace{-9mm}
  \includegraphics[width=0.98\textwidth]{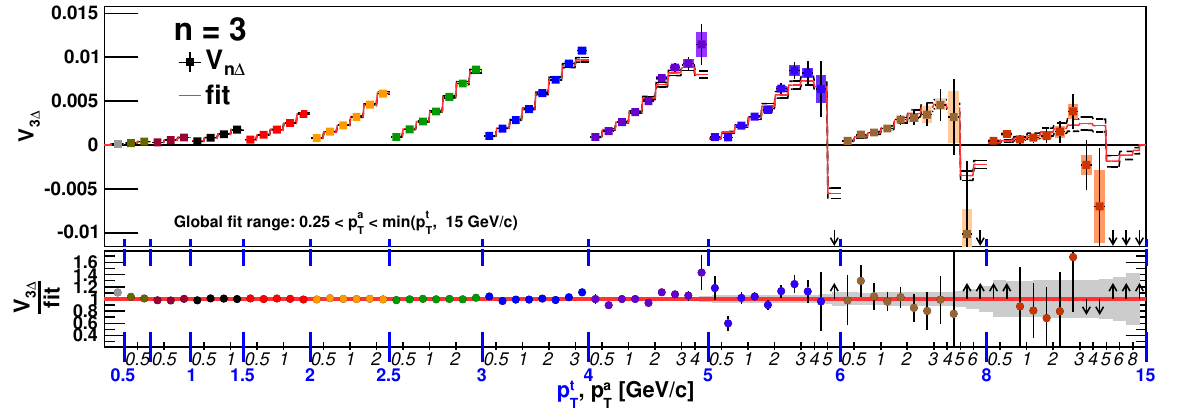} \\
  \vspace{-9mm}
  \includegraphics[width=0.98\textwidth]{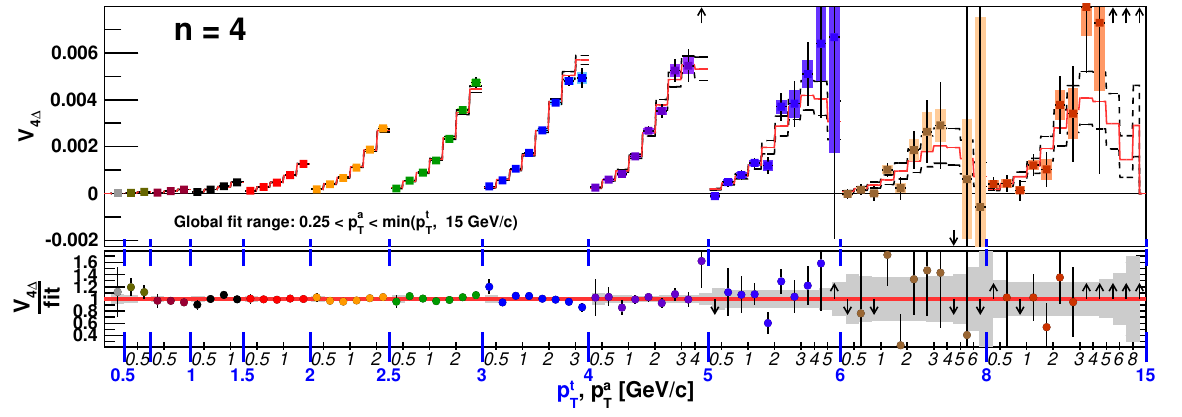} \\
  \vspace{-9mm}
  \includegraphics[width=0.98\textwidth]{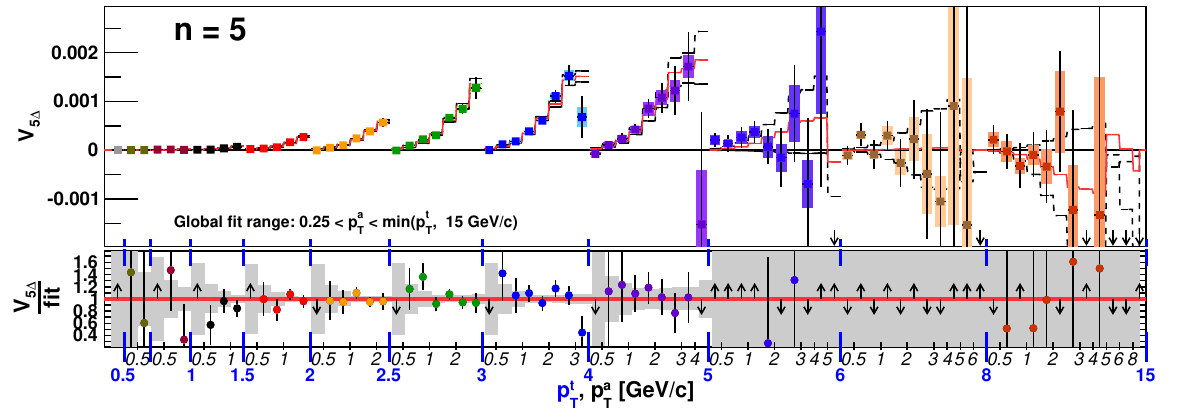}\\
  \vspace{-4mm}
  \caption[Global vn results]{\label{fig:gf} (Color online) Global fit
    examples in $0$--$10$\% central events for $n=2,3,4$ and $5$. The
    measured $\vnd$ coefficients are plotted on an interleaved $\ptt$,
    $\pta$ axis in the upper panels, and the global fit function
    (\Eq{eq:fac}) is shown as the red curves. The global fit
    systematic uncertainty is represented by dashed lines. The lower
    section of each panel shows the ratio of the data to the fit, and
    the shaded bands represent the systematic uncertainty propagated
    to the ratio. In all cases, off-scale points are indicated with
    arrows. }
\end{figure}

\Equation{eq:fac} is tested by applying a global fit to the $\vnd$
data points over all $\ptt$ and $\pta$ bins simultaneously. This is
done separately at each order in $n$ and for each centrality class. An
example from the $0$--$10$\% most central event class is shown in
\Fig{fig:gf}, where the $\vnd$ points for $n=2$ to 5 are plotted (in
separate panels) on a single $\ptt$, $\pta$ axis as indicated. The
global fit function depends on a set of $N$ unconstrained and
independent parameters, where $N$ is the number of $\ptt$ (or $\pta$)
bins. The parameters are $\vngf(\pt)$, with the fit generating the
product $\vngf(\ptt) \times \vngf(\pta)$ that minimizes the total
$\chi^2$ for all $\vnd$ points.

The sources of systematic uncertainty of $\vnd$ are those that cause
$\df$-dependent variation on $\CF$. Factors affecting overall yields
such as single-particle inefficiency cancel in the ratio
of~\Eq{eq:cfdef}, and do not generate uncertainty in $\CF$.
Table~\ref{tab:vndsys} shows the different contributions to the
systematic uncertainty of $\vnd$, and Table~\ref{tab:vndsys_typ} lists
typical magnitudes of these uncertainties for a few representative
centrality classes.

The event mixing uncertainty (denoted as ``a'' in
Table~\ref{tab:vndsys}) accounts for biases due to imperfect matching
of event multiplicity and collision vertex position, as well as for finite
mixed-event statistics. This uncertainty changes with $\ptt$,
$\pta$, and centrality. It is evaluated by comparing the $n\leq5$
Fourier sum from $\CF$ with that from
$N_{\mathrm{same}}(\df)$. 
The uncertainty from (a) is depicted by grey bars on the points
in $\CF$ in~\Fig{fig:cf_low} and ~\Fig{fig:cf_high}. Due to
fluctuations in the mixed-event distribution, uncertainty (a) tends to
scale with the $\vnd$ statistical error, as shown in the table.

The remainder of the systematic uncertainties are not assigned to
$\CF$, but rather to each $\vnd$ directly, where their influence on
$\vnd$ is more clearly defined. The uncertainty from centrality
determination (b) accounts for the resolution and efficiency of the
detector used for multiplicity measurements, as well as any biases
related to its $\eta$ acceptance. This uncertainty is globally
correlated in centrality. It was studied by conducting the full
analysis with the SPD as an alternative centrality estimator, since it
has different systematic uncertainties and covers a different
pseudorapidity range than the $\VZERO$ detectors. The results were
found to agree within 1\%.

Uncertainty from tracking and momentum resolution (c) was evaluated on
$\CF$ using different track selection criteria. Slightly larger
correlation strength is obtained for more restrictive track selection
(at the expense of statistical loss), and the difference was found to
grow with $\pta$ by roughly 1\% per GeV/$c$.

Additional uncertainty is introduced by (d) the finite $\df$ bin
width, which was estimated by comparing the RMS bin width to the
$n^{\mathrm{th}}$ harmonic
scale, 
and (e) the precision of the extraction (\Eq{eq:vnd}). The latter was
estimated by calculating $\langle \sin (n \df) \rangle$, which is
independent of $n$, and should vanish by symmetry. The residual finite
values are used to gauge the corresponding $\vnd$ uncertainty. Because
the amplitude of the $\vnd$ harmonics tends to diminish with
increasing $n$, both uncertainties (d) and (e) are small for $n<6$ but
become comparable to $\vnd$ at higher $n$. Effects (a)-(e) are all
combined in quadrature to produce the $\vnd$ systematic uncertainties,
which are depicted as the solid colored bars on the points in
\Fig{fig:VnDelta} and \Fig{fig:gf}. Finally, the uncertainty (f) is
included in the quadrature sum with (a)-(e) for $\voned$ only. 

\clearpage

\begin{table}[bht] \centering
  \begin{tabular}{llll}
    \hline
    Contribution & Magnitude \\
    \hline
    (a) Event mixing   & $20$-$30\% \, \sigma_{\mathrm{stat}}$  \\
    (b) Centrality determination & 1\% $\vnd$ \\
    (c) Track selection, $\pt$ resolution & $1\% \vnd \times \langle \pta \rangle$  \\
    (d) $\df$ bin width &  $(0.8 n)\% \vnd$ \\
    (e) $\vnd$ extraction  & $<$10\% $\vnd$ ($n$$<$6); 10-30\% \vnd ($n$$\geq$6)\\
    (f) $\vnd$ ($n$=1 only)  & $10\% \vnd \times \langle \ptt \rangle \langle \pta \rangle$\\
    \hline
  \end{tabular}
  \caption{\label{tab:vndsys} Systematic uncertainties on $\vnd$.}
\end{table}

\begin{table}[bht] \centering
\begin{tabular}{rlrrrrrrrr}
 $n$  &  Centrality  &  $\langle \vnd \rangle$  &
 $\langle \sigma_{sys (tot)} \rangle$   & $\sigma_a$  &  $\sigma_b$  & $\sigma_c$  & $\sigma_d$  & $\sigma_e$  & $\sigma_f$  \\
  &   & $(\times 10^{-3})$  & $ (\times 10^{-3})$   &   &  &  $ (\times 10^{-3})$  &  & &  \\

\hline
 1  &  0-10\%   &    4.2  &      13  &  0.57  &   0.042  &   0.042  &  0.033  &    13  &  0.42  \\
 1  &  20-30\%  &     11  &     3.7  &   0.6  &    0.11  &    0.11  &  0.089  &   3.5  &   1.1  \\
 1  &  40-50\%  &     23  &     2.3  &  0.38  &    0.23  &    0.23  &   0.18  &  0.31  &   2.3  \\
 2  &  0-10\%   &     12  &       7  &  0.23  &    0.12  &    0.12  &   0.19  &     7  &     0  \\
 2  &  20-30\%  &     31  &       1  &  0.46  &    0.31  &    0.31  &    0.5  &   0.6  &     0  \\
 2  &  40-50\%  &     43  &     6.9  &   0.4  &    0.43  &    0.43  &   0.69  &   6.8  &     0  \\
 3  &  0-10\%   &   0.82  &      18  &  0.29  &  0.0082  &  0.0082  &   0.02  &    18  &     0  \\
 3  &  20-30\%  &    2.1  &     1.4  &  0.42  &   0.021  &   0.021  &   0.05  &   1.4  &     0  \\
 3  &  40-50\%  &     10  &     4.2  &  0.42  &     0.1  &     0.1  &   0.25  &   4.2  &     0  \\
 4  &  0-10\%   &    2.6  &     5.4  &  0.44  &   0.026  &   0.026  &  0.083  &   5.4  &     0  \\
 4  &  20-30\%  &    7.2  &    0.45  &  0.27  &   0.072  &   0.072  &   0.23  &  0.26  &     0  \\
 4  &  40-50\%  &     11  &     2.6  &  0.47  &    0.11  &    0.11  &   0.35  &   2.6  &     0  \\
 5  &  0-10\%   &    2.6  &     6.2  &  0.32  &   0.026  &   0.026  &    0.1  &   6.1  &     0  \\
 5  &  20-30\%  &    2.8  &     3.2  &  0.17  &   0.028  &   0.028  &   0.11  &   3.2  &     0  \\
 5  &  40-50\%  &    6.8  &     0.6  &  0.28  &
 0.068  &   0.068  &   0.27  &  0.45  &     0  \\
\hline
\end{tabular}
\caption{\label{tab:vndsys_typ} Typical values of $\vnd$ systematic
  uncertainties.}
\end{table}

To evaluate the systematic uncertainty, the global fit procedure is
performed three times for each $n$ and centrality bin: once on the
measured $\vnd$ points (leading to the red curves in \Fig{fig:gf}), and
once on the upper and lower bounds of the systematic error
bars~(resulting in black dashed curves). The $\vngf$ systematic error
is then assigned as half the difference. The resulting uncertainties
are shown as open boxes in \Fig{fig:vnglobal} and \Fig{fig:v1}, which
are discussed in the following sections.

\section{Global fit results} \label{sec:vngf} In the $n=2$ case
(\Fig{fig:gf}, top), the fit agrees well with the data points
at low $\ptt$ and \pta, but diverges with increasing $\pta$ for each
$\ptt$ interval. Where disagreement occurs, the fit is systematically
lower than the points. In contrast, for $n=3$, the fit
does not follow the points that drop sharply to negative values at the
highest momenta. This is also observed for $n=5$, though with poorer
statistical precision.

The global fit is driven primarily by lower particle $\pt$, where the
smaller statistical uncertainties provide a stronger constraint for
$\chi^2$ minimization.  The disagreement between data and the fit,
where $\ptt$ and $\pta$ are both large, points to the breakdown of the
factorization hypothesis; see also \Fig{fig:cf_high} and the
accompanying discussion.

The factorization hypothesis appears to hold for $n\ge2$ at low $\pta$
($\lesssim 2$ GeV/$c$) even for the highest $\ptt$ bins. The $\vnd$
values for these cases are small relative to those measured at higher
$\pta$, and remain constant or even decrease in magnitude as $\ptt$ is
increased above 3-4 GeV/$c$. $V_{2\Delta}$ dominates over the other
coefficients, and the $n>3$ terms are not significantly greater than
zero. This stands in contrast to the high-$\ptt$, high-$\pta$ case,
where it was demonstrated in \Fig{fig:cf_high} that dijet correlations
require significant high-order Fourier harmonics to describe the
narrow recoil jet peak. 

\begin{figure}[tbh] \centering 
  \includegraphics[width=0.98\textwidth]{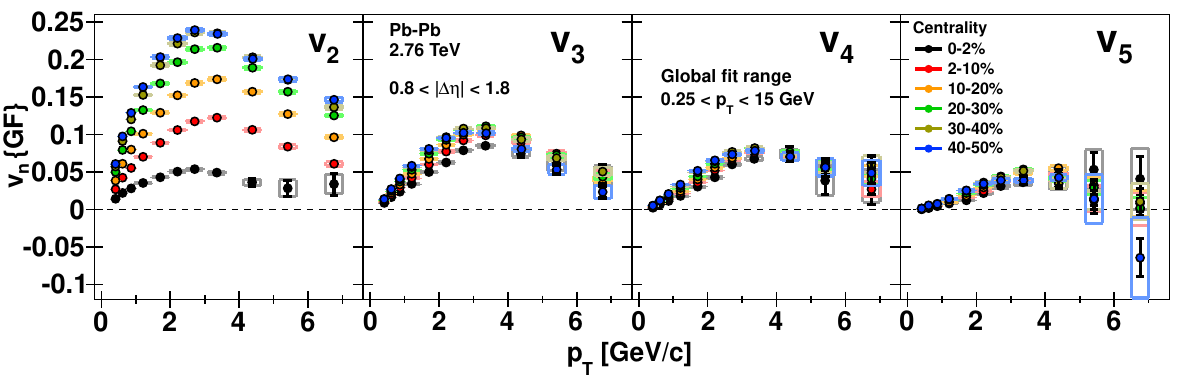}
  \caption[Global vn results]{\label{fig:vnglobal}(Color online) The
    global-fit parameters, $\vngf$, for $2 \leq n \leq 5$. Statistical
    uncertainties are represented by error bars on the points, while
    systematic uncertainty is depicted by open rectangles.}
\end{figure}

The parameters of the global fit are the best-fit $\vngf$ values as a
function of $\pt$, which can be interpreted as the coefficients of
\Eq{eq:vn}. The results of the global fit for $2 \leq n \leq 5$,
denoted $\vngf$, are shown in \Fig{fig:vnglobal} for several
centrality selections. We note that the global fit converges to either
positive or negative $\vngf$ parameters, depending on the starting
point of the fitting routine. The two solutions are equal in magnitude
and goodness-of-fit. The positive curves are chosen by convention as
shown in \Fig{fig:vnglobal}. In the $0$--$2$\% most central data,
$v_3\{ GF \}$ ($v_4\{ GF \}$) rises with $\pt$ relative to $v_2\{ GF
\}$ and in fact becomes larger than $v_2\{ GF \}$ at approximately 1.5
(2.5) GeV/$c$. $v_2\{ GF \}$ reaches a maximum value near 2.5 GeV/$c$,
whereas the higher harmonics peak at higher $\pt$. These data are in
good agreement with recent two-particle anisotropic flow
measurements~\cite{Aamodt:2011vk} at the same collision energy, which
included a pseudorapidity gap of $|\de|>1.0$.

For $2 \leq n \leq 5$, the results are not strongly sensitive
to the upper $\pta$ limit included in the global fit. The global fit
was performed not only over the full momentum range (as shown in
\Fig{fig:vnglobal}), but also with the restriction to $\vnd$ points
with $\pta < 2.5$ GeV/$c$. The outcome was found to be identical to
the full fit within one standard deviation. This again reflects the
weighting by the steeply-falling particle momentum distribution,
indicating that a relatively small number of energetic particles does
not strongly bias the event anisotropy, as calculated by the global
fit.

\begin{figure}[tbh] \centering 
  \includegraphics[width=0.4\textwidth]{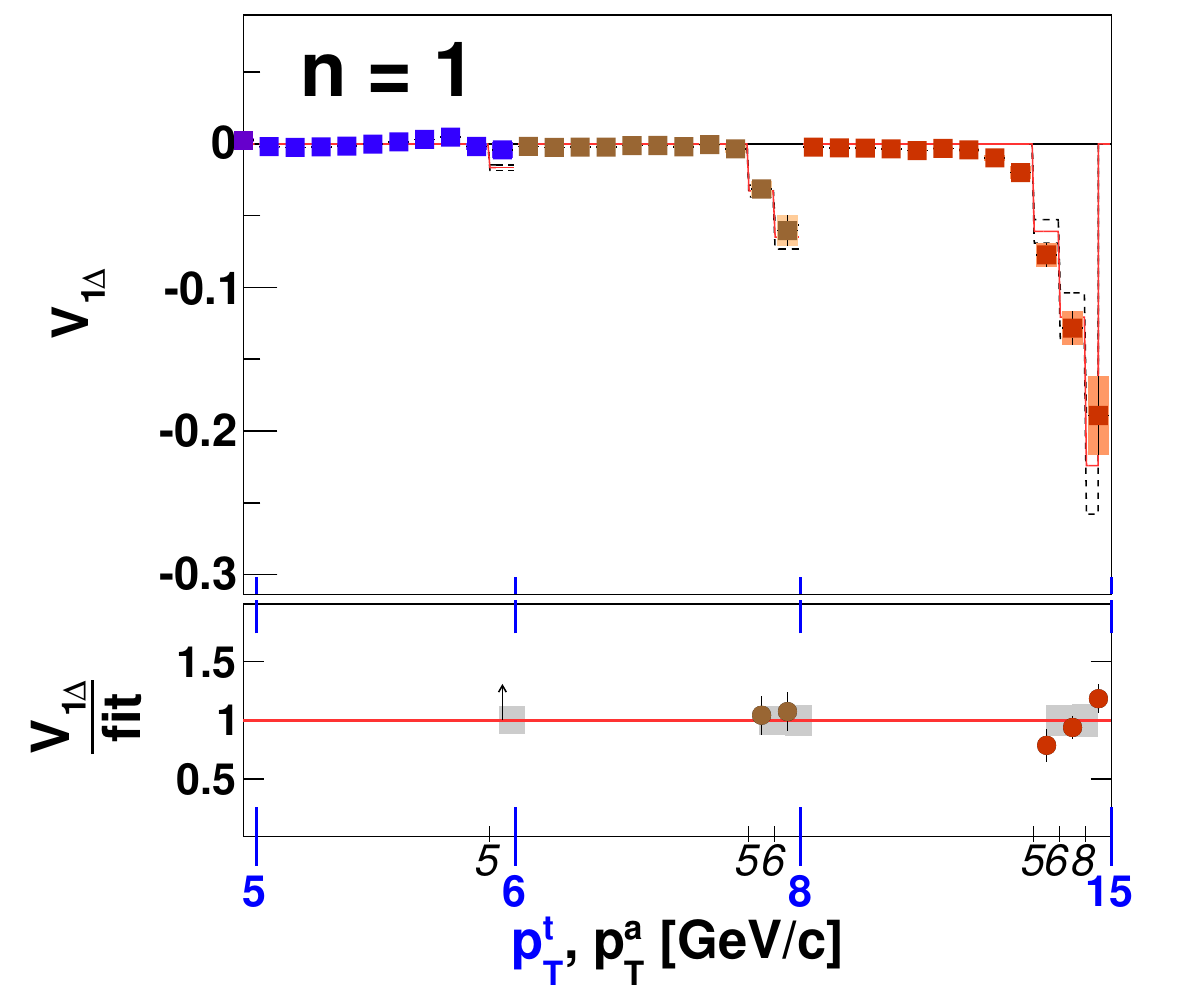}
  \qquad
  \includegraphics[width=0.4\textwidth]{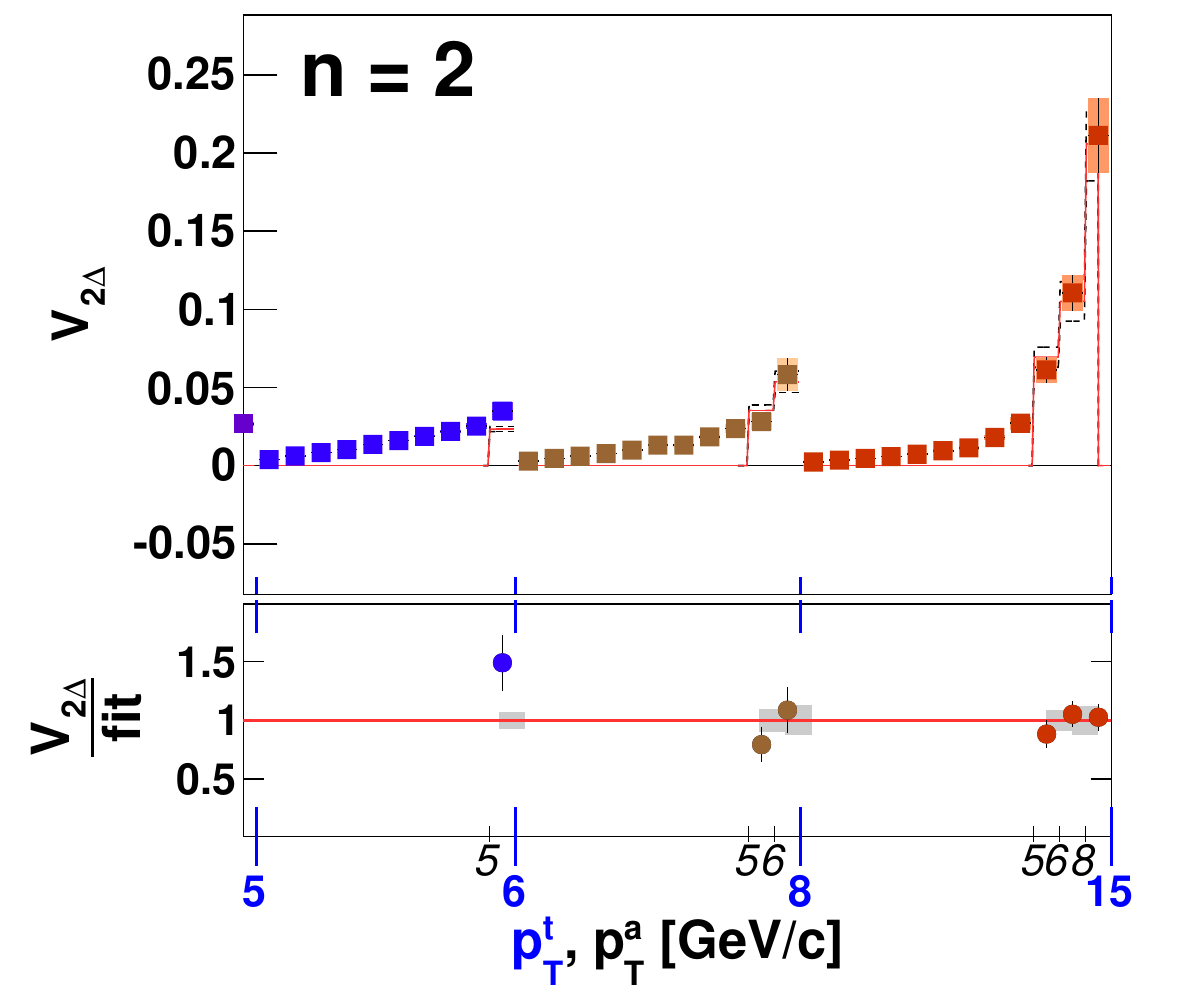} \\
  \vspace{-9mm}
  \includegraphics[width=0.4\textwidth]{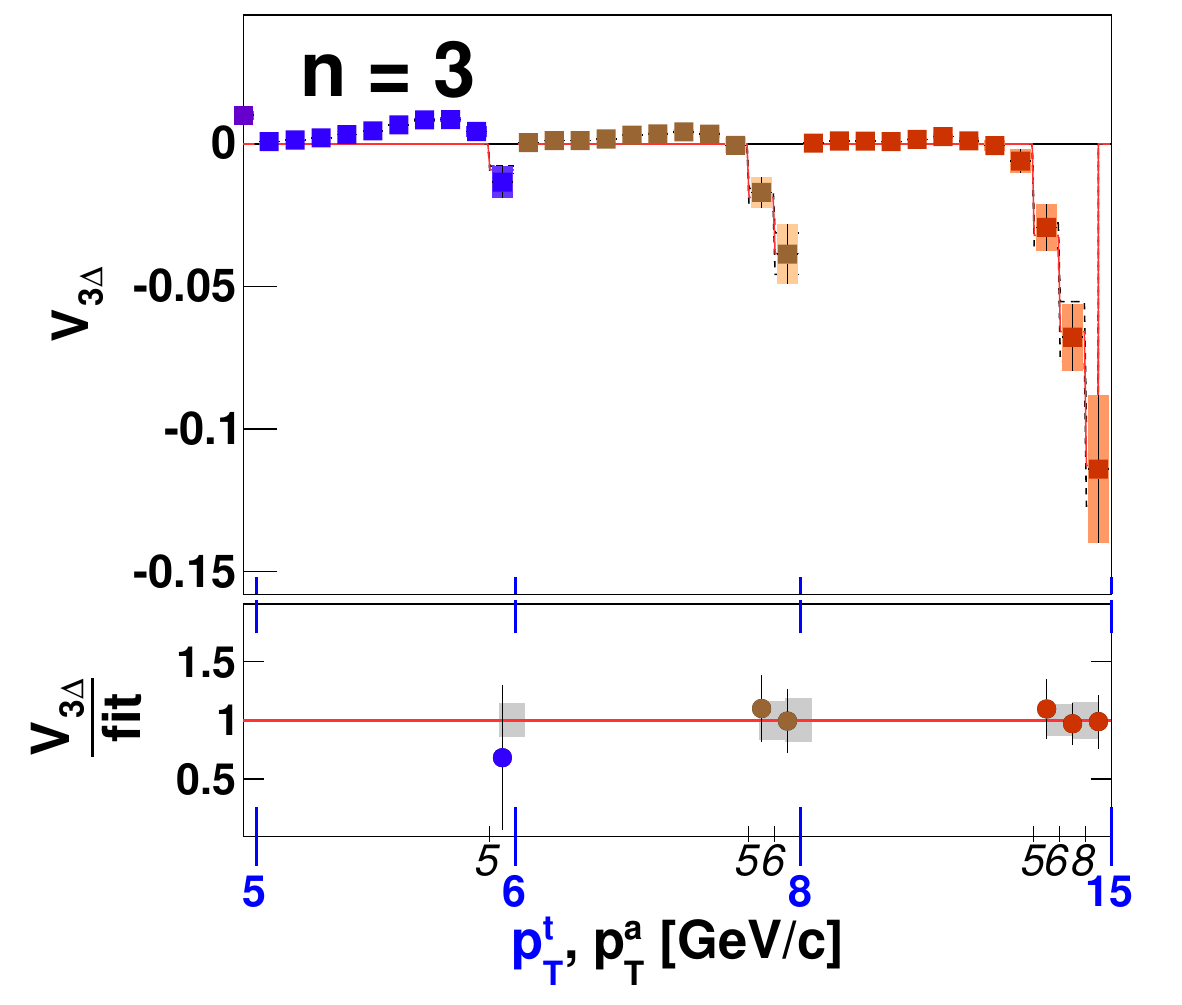}
  \qquad
  \includegraphics[width=0.4\textwidth]{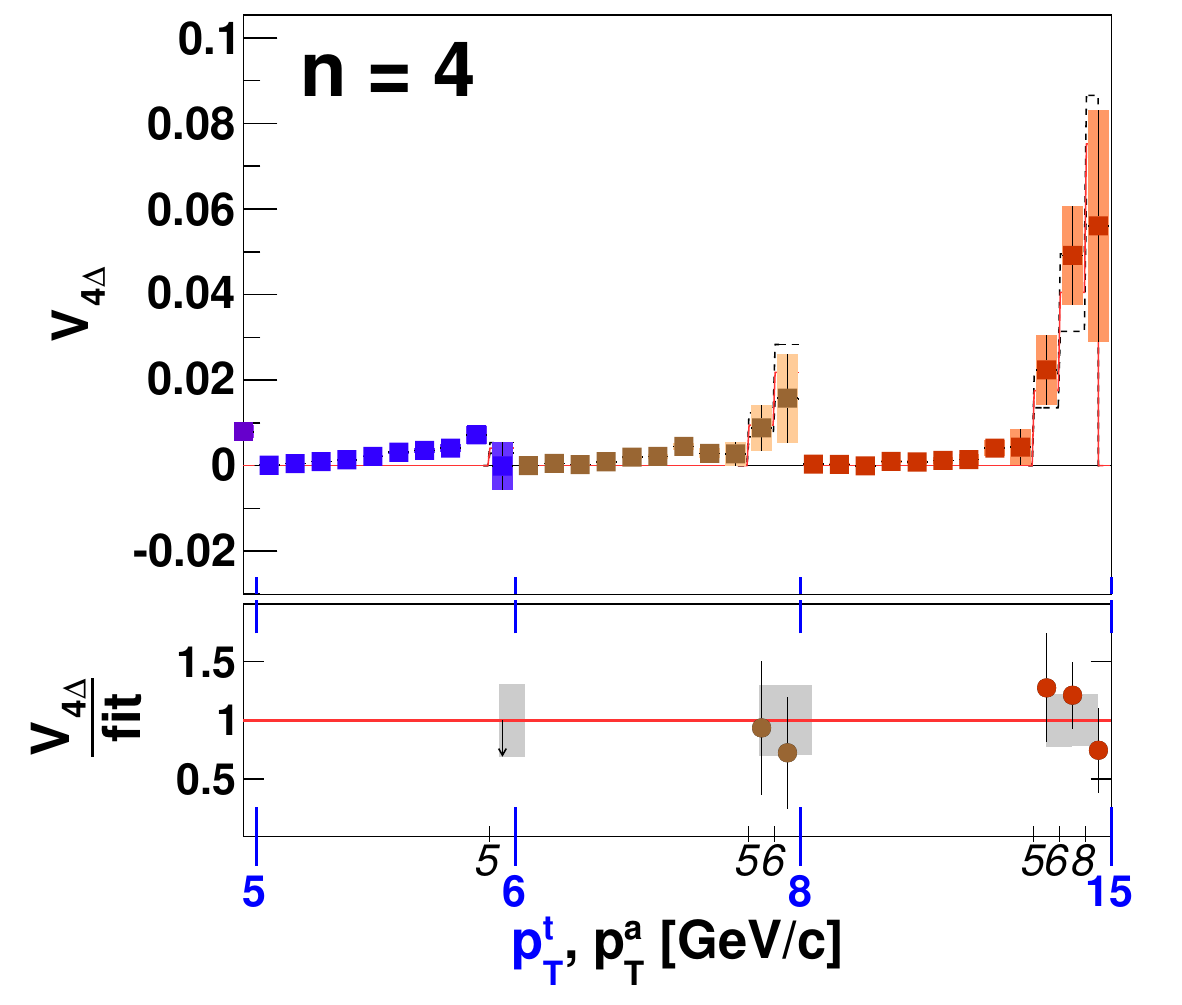} \\
  \vspace{-4mm}
  \caption[Global fit - high pt]{\label{fig:gfhi} (Color online)
    High-$\pt$ fit examples in $0$--$20$\% central events for $n=1$ to
    $4$. Although all datapoints are shown for $\ptt > 5$ GeV/$c$, the
    fit range includes only the six points with $ \pta > 5$ GeV/$c$.}
\end{figure}

If the global fit is applied to $\vnd$ points exclusively at large
particle momenta, factorization behavior can be tested for
correlations that are predominantly jet-induced. An example is shown
in figure \Fig{fig:gfhi}, where the global fit has been applied to
$\vnd$ points within $5 < \pta < 15$ GeV/$c$. In this case, there are
six $\vnd$ datapoints fitted, and three fit parameters, which are
$\vngf$ at 5-6, 6-8, and 8-15 GeV/$c$. An approximate factorization is
observed over this range. The agreement between fit and data for the
lowest fitted datapoint (at 5-6 GeV/$c$) is rather poor, indicating
that the correlations there are in a transitional region that is less
jet-dominated than at higher $\pt$.

The parameters from the high-$\pt$ global fit can be plotted, just as
was done in \Fig{fig:vnglobal}, to demonstrate their $\pt$ and
centrality dependence. However, the sign definition of $\vn$ becomes
problematic in the case where $\vn(\ptt)$ and $\vn(\pta)$ have the
same sign, but $\vnd < 0$. In this case, the $\vn$ coefficients are
represented to be positive as a matter of convention.

\begin{figure}[tbh] \centering 
  \includegraphics[width=0.98\textwidth]{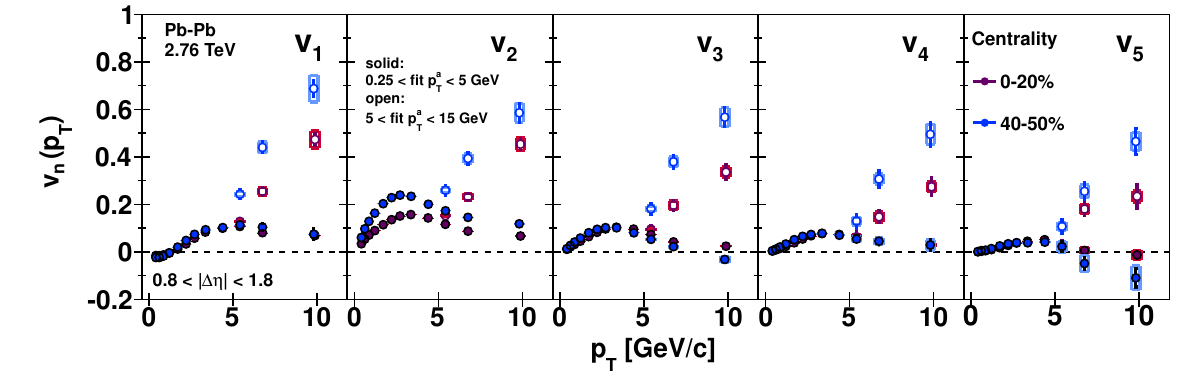}
  \caption[Global vn results]{\label{fig:vnsplit} (Color
    online) The global-fit parameters, $\vngf$, for $1 \leq n \leq 5$
    as obtained using restricted $\pta$ fit ranges at two different
    centralities. The solid (open) points represent fits using only
    $0.25 < \pta < 5$ ($5 < \pta < 15$) GeV/$c$. The open points
    represent the magnitude of $\vngf$ from high-$\ptt$, high-$\pta$
    long-range correlations. Statistical uncertainties are represented
    by error bars on the points, while systematic uncertainty is
    depicted by open rectangles.}
\end{figure}

The fit results from these high-$\ptt$, high-$\pta$ long-range
correlations are shown as open points in \Fig{fig:vnsplit}. The clear
deviation between the two different sets of points demonstrates that
it is not possible for a single-valued set of $\vn(\pt)$ points to
simultaneously describe both low-$\pta$ and high-$\pta$ pair
anisotropy.

\begin{figure}[tbh] \centering
  \includegraphics[width=0.98\textwidth]{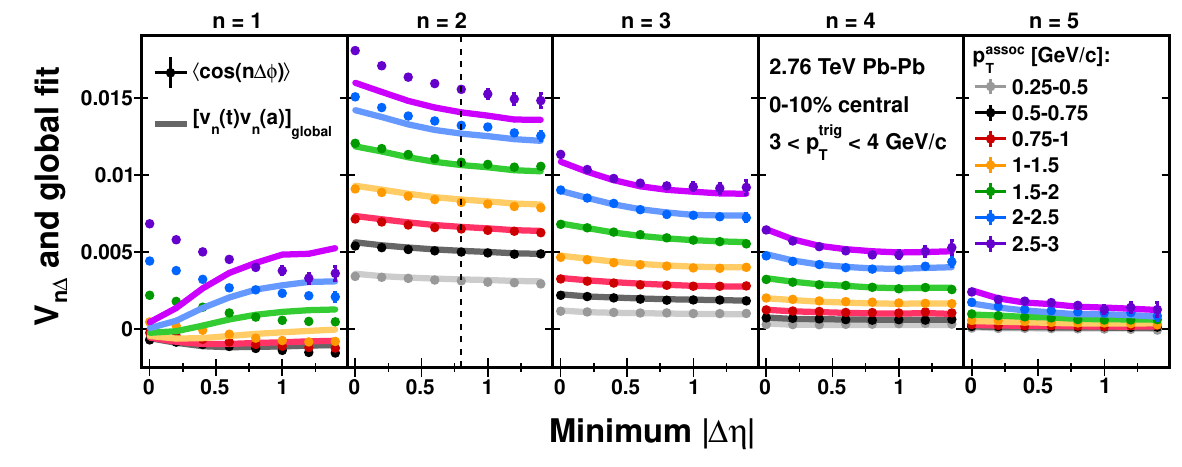}
  \caption[V2DeltaHi]{\label{fig:etagap} (Color online) $\vnd$ values
    from $0$--$10$\% central Pb--Pb collisions (points) and global fit
    results (solid lines) for $3.0 < \ptt < 4.0$~GeV/$c$ as a function of the
    minimum $|\de|$ separation for a selection of $\pta$ bins. For
    clarity, points are shown with statistical error bars only. For
    reference, a dashed line (drawn only in the $n=2$ panel) indicates
    the $|\de|_{\mathrm{min}} = 0.8$ requirement applied throughout
    this analysis.}
\end{figure}

It is instructive to study the dependence of the $\vnd$ values on the
minimum $|\de|$ separation in order to observe the influence of the
near-side peak. This is shown in \Fig{fig:etagap}. The $\vnd$ values
rise as the pseudorapidity gap is reduced and a larger portion of the
near-side peak is included in the correlations. At $\ptt > 3$--$4$
GeV/$c$, the peak is narrow and the curves are fairly flat at $|\de| >
0.5$. For the 3--4 GeV/$c$ range shown in the figure, there is a
discernible contribution from the near-side peak, but the difference
does not exceed a few percent at $|\de| > 0.8$. 

For the first harmonic, the disagreement grows significantly as
  $|\de|$ is decreased, while the higher harmonics exhibit a much
  lower sensitivity to $|\de|_{min}$.
  Even if a large $\eta$ gap is applied, however, \Fig{fig:etagap}
  indicates that an accurate global description of $\voned$ is still
  not obtained, even at these low to intermediate $\pt$ values.

\begin{figure}[tbh] \centering
  \includegraphics[width=0.98\textwidth]{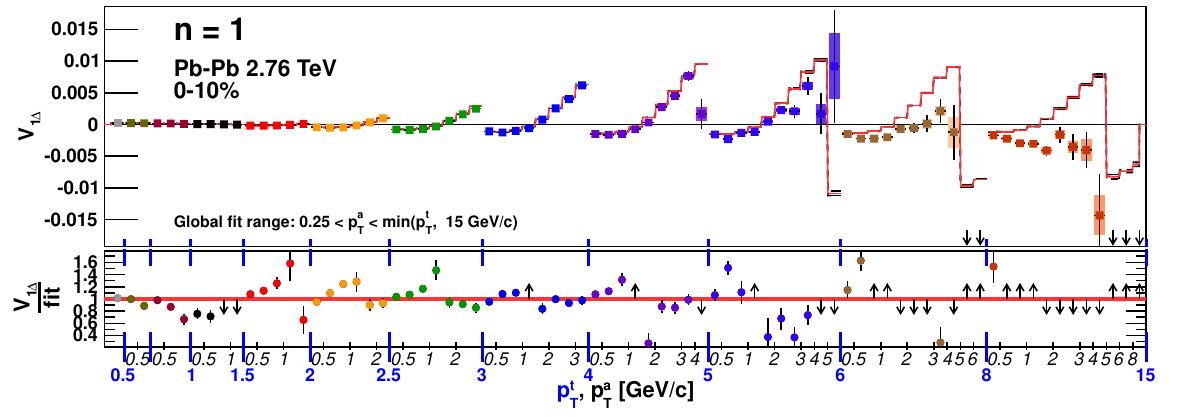} \\
  \vspace{-9mm}
  \includegraphics[width=0.98\textwidth]{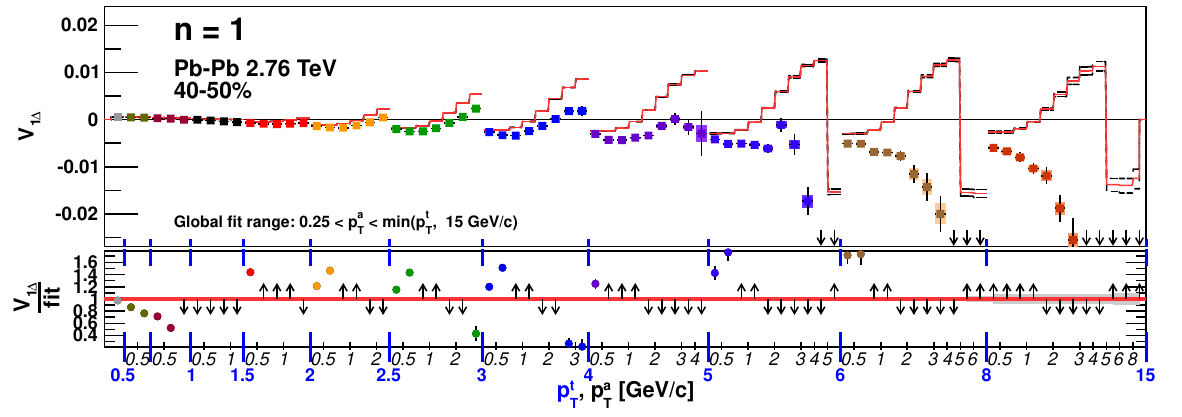}
  \caption[Global v1 results]{\label{fig:v1gf} (Color online) Global
    fit examples in $0$--$10$\% central (top) and $40$--$50$\%
    central events (bottom) for $n=1$. The uncertainties are
    represented in the same way as for \Fig{fig:gf}.}
\end{figure}
\begin{figure}[hbt] \centering
  \includegraphics[width=0.48\textwidth]{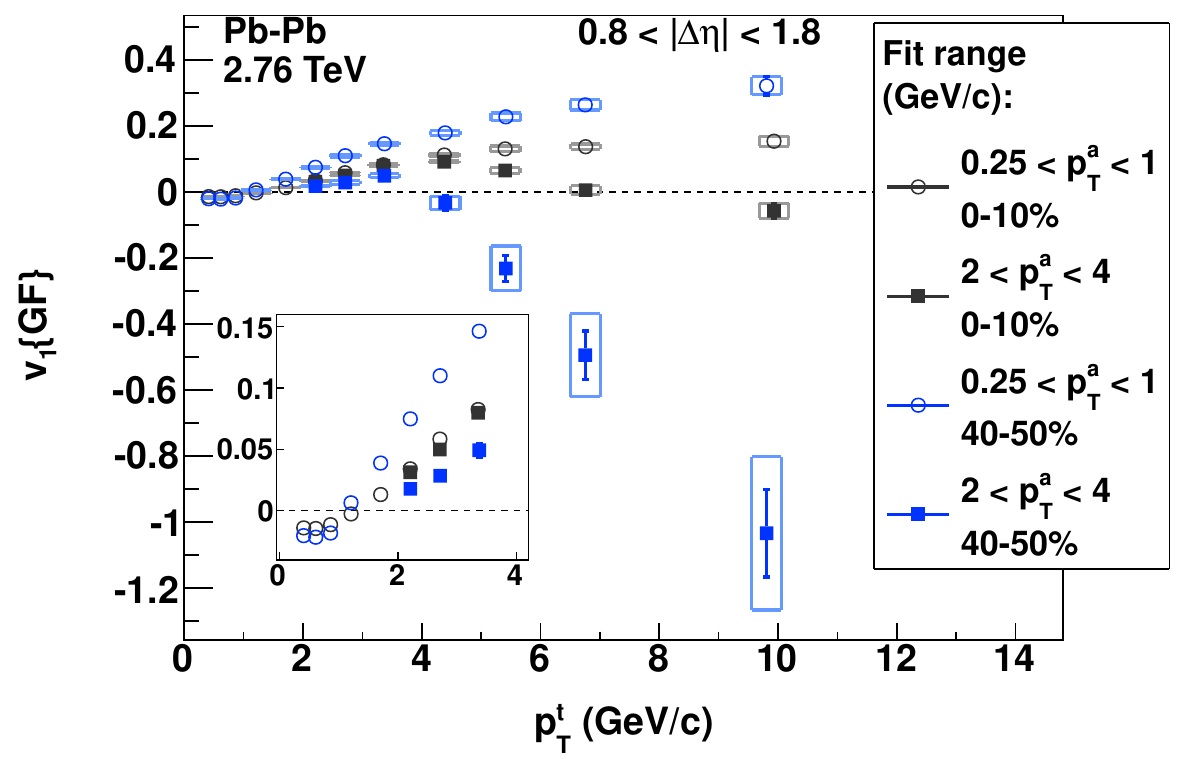}
  \includegraphics[width=0.48\textwidth]{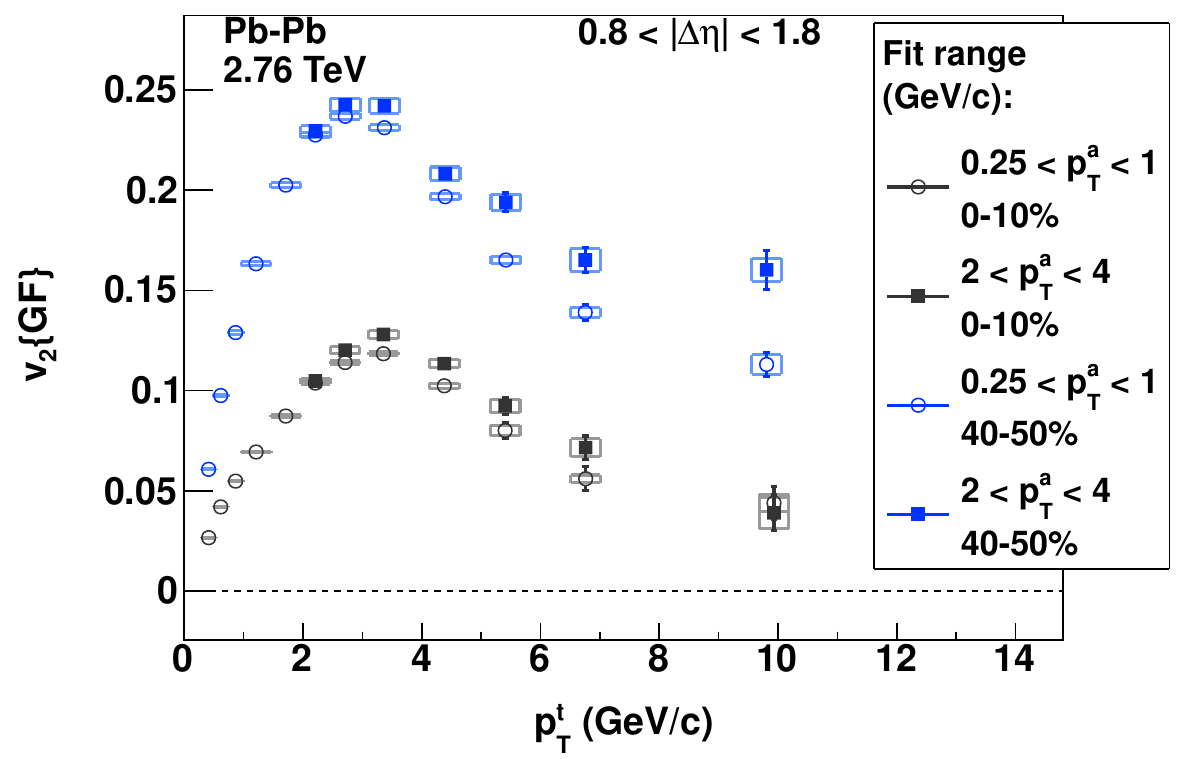}
  \caption[v1]{\label{fig:v1} (Color online) $v_1\{GF\}$ (left) and
    $v_2\{GF\}$ (right) as obtained using restricted $\pta$ fit ranges
    at two different centralities. The open circles (solid squares)
    represent fits using only $0.25 < \pta < 1$ ($2 < \pta < 4$)
    GeV/$c$. For the more central $v_2$ points, the two different fit
    ranges lead to a similar curve, indicating an approximate
    factorization. In contrast, a divergence with rising $\ptt$ is
    observed for $v_1\{GF\}$. For both $n=1$ and $n=2$, the divergence
    is enhanced in more peripheral centrality classes.  }
\end{figure}

This behavior is representative of a general lack of consistent
$V_{1\Delta}$ factorization, as demonstrated in \Fig{fig:v1gf} where
the global fit for $n=1$ is shown for two centrality ranges.
However, a reasonable fit is obtained if the $\pta$ range is
  restricted to smaller intervals. \Fig{fig:v1} shows the result of
  performing the global fit to $v_1$ (left) and $v_2$ (right) over
  $0.25 < \pta < 1.0$ and $2<\pta<4$ GeV/$c$ separately at two
  different centralities. In the case of $v_1$, a divergence occurs
  between the results obtained from the two different $\pta$ bins,
  which is more prominent in mid-central collisions. For 0-20\% $v_2$,
  however, fits using the two different $\pta$ intervals lead to
  approximately the same curve, supporting the observation that $v_2$
  factorizes: on average over events in this centrality category, a
  unique symmetry plane exists for the majority of all particles below
  4 GeV/$c$. The systematic increase of the higher $\pta$ fit compared
  to $\pta < 1$ is likely from nonflow contributions on the away side,
  which are larger in the more peripheral centralities because of
  reduced quenching effects. Thus the observed patterns follow the
  expected trends with $\pta$ and centrality.

The breakdown of factorization for $\voned$ does not imply that
there is no real collective $v_1$ since the collective part may not be
the dominant contribution to $\voned$.  It is therefore interesting to
note that at low $\pt$, the best-fit $v_1\{GF\}$ values become
negative, as observed in hydrodynamic simulations with fluctuating
initial conditions~\cite{Gardim:2011qn}.  Although those calculations
were for Au+Au collisions at 200 GeV, qualitatively similar results
have been obtained at 2.76 TeV~\cite{luzum_privcomm}.  Estimation of
the effect of momentum conservation as a correction to the
coefficients prior to the global fit is currently under
investigation. In~\cite{Gardim:2011qn}, such a correction amounted to
a change in $v_1$ of about $0.01$-$0.02$. In this analysis we have
included a systematic uncertainty of $10\% \, \langle \pta \rangle$ in
$\voned$ to account for the bias resulting from the neglect of this
correction, and the uncertainty is propagated to $v_1\{GF\}$ in the
same fashion as for $n > 1$.
Further studies will be required to unambiguously extract the collective 
part of $\voned$.

\section{Summary} \label{sec:summary} The shape evolution of triggered
pair distributions was investigated quantitatively using a discrete
Fourier decomposition. In the bulk-dominated $\pt$ regime, a distinct
near-side ridge and a doubly-peaked away-side structure are observed
in the most central events, both persisting to a large relative
pseudorapidity interval between trigger and associated
particles. These features are represented in Fourier spectra by
harmonic amplitudes, both even and odd, which are finite in magnitude
up to approximately $n=5$. These pair anisotropies are found to
approximately factorize into single-particle harmonic
coefficients for $\pta < 4$ GeV/$c$, with the notable exception
  of $\voned$. This factorization is consistent with expectations
from collective response to anisotropic initial conditions, which
provides a complete and self-consistent picture explaining the
observed features without invocation of dynamical mechanisms such as
Mach shock waves~\cite{CasalderreySolana:2004qm}.

The data also suggest that at low $\pt$ (below approximately 3
GeV/$c$), any contribution from the away-side jet is constrained to be
relatively small. In contrast, for associated $\pt$ greater than
$4$--$6$ GeV/$c$, the long-range correlation appears dominated by a
large peak from the recoil jet. In this regime, when both particles
are at high momenta, the anisotropy does not follow the
$\pt$-dependent pattern followed by particle pairs at lower $\pta$.
The global fit technique provides a means of identifying transitions
in the momentum and centrality dependence of correlations with respect
to symmetry planes.  Within the bulk-dominated region, the measurement
of all significant harmonics provides the possibility to constrain the
geometry of the fluctuating initial state and further understand the
nuclear medium through its collective response.


\ifpreprint
\iffull
\newenvironment{acknowledgement}{\relax}{\relax}
\begin{acknowledgement}
\section*{Acknowledgements}
The ALICE collaboration would like to thank all its engineers and technicians for their invaluable contributions to the construction of the experiment and the CERN accelerator teams for the outstanding performance of the LHC complex.
\\
The ALICE collaboration acknowledges the following funding agencies for their support in building and
running the ALICE detector:
 \\
Department of Science and Technology, South Africa;
 \\
Calouste Gulbenkian Foundation from Lisbon and Swiss Fonds Kidagan, Armenia;
 \\
Conselho Nacional de Desenvolvimento Cient\'{\i}fico e Tecnol\'{o}gico (CNPq), Financiadora de Estudos e Projetos (FINEP),
Funda\c{c}\~{a}o de Amparo \`{a} Pesquisa do Estado de S\~{a}o Paulo (FAPESP);
 \\
National Natural Science Foundation of China (NSFC), the Chinese Ministry of Education (CMOE)
and the Ministry of Science and Technology of China (MSTC);
 \\
Ministry of Education and Youth of the Czech Republic;
 \\
Danish Natural Science Research Council, the Carlsberg Foundation and the Danish National Research Foundation;
 \\
The European Research Council under the European Community's Seventh Framework Programme;
 \\
Helsinki Institute of Physics and the Academy of Finland;
 \\
French CNRS-IN2P3, the `Region Pays de Loire', `Region Alsace', `Region Auvergne' and CEA, France;
 \\
German BMBF and the Helmholtz Association;
\\
General Secretariat for Research and Technology, Ministry of
Development, Greece;
\\
Hungarian OTKA and National Office for Research and Technology (NKTH);
 \\
Department of Atomic Energy and Department of Science and Technology of the Government of India;
 \\
Istituto Nazionale di Fisica Nucleare (INFN) of Italy;
 \\
MEXT Grant-in-Aid for Specially Promoted Research, Ja\-pan;
 \\
Joint Institute for Nuclear Research, Dubna;
 \\
National Research Foundation of Korea (NRF);
 \\
CONACYT, DGAPA, M\'{e}xico, ALFA-EC and the HELEN Program (High-Energy physics Latin-American--European Network);
 \\
Stichting voor Fundamenteel Onderzoek der Materie (FOM) and the Nederlandse Organisatie voor Wetenschappelijk Onderzoek (NWO), Netherlands;
 \\
Research Council of Norway (NFR);
 \\
Polish Ministry of Science and Higher Education;
 \\
National Authority for Scientific Research - NASR (Autoritatea Na\c{t}ional\u{a} pentru Cercetare \c{S}tiin\c{t}ific\u{a} - ANCS);
 \\
Federal Agency of Science of the Ministry of Education and Science of Russian Federation, International Science and
Technology Center, Russian Academy of Sciences, Russian Federal Agency of Atomic Energy, Russian Federal Agency for Science and Innovations and CERN-INTAS;
 \\
Ministry of Education of Slovakia;
 \\
CIEMAT, EELA, Ministerio de Educaci\'{o}n y Ciencia of Spain, Xunta de Galicia (Conseller\'{\i}a de Educaci\'{o}n),
CEA\-DEN, Cubaenerg\'{\i}a, Cuba, and IAEA (International Atomic Energy Agency);
 \\
Swedish Reseach Council (VR) and Knut $\&$ Alice Wallenberg Foundation (KAW);
 \\
Ukraine Ministry of Education and Science;
 \\
United Kingdom Science and Technology Facilities Council (STFC);
 \\
The United States Department of Energy, the United States National
Science Foundation, the State of Texas, and the State of Ohio.
\end{acknowledgement}
\ifbibtex
\bibliographystyle{unsrtnat}
\bibliography{biblio}{}
\else

\fi
\newpage
\appendix
\section{The ALICE Collaboration}
\label{app:collab}

\begingroup
\small
\begin{flushleft}
K.~Aamodt\Irefn{org1121}\And
B.~Abelev\Irefn{org1234}\And
A.~Abrahantes~Quintana\Irefn{org1197}\And
D.~Adamov\'{a}\Irefn{org1283}\And
A.M.~Adare\Irefn{org1260}\And
M.M.~Aggarwal\Irefn{org1157}\And
G.~Aglieri~Rinella\Irefn{org1192}\And
A.G.~Agocs\Irefn{org1143}\And
A.~Agostinelli\Irefn{org1132}\And
S.~Aguilar~Salazar\Irefn{org1247}
% ...
\renewcommand\labelenumi{\textsuperscript{\theenumi}~}
\section*{Affiliation notes}
\renewcommand\theenumi{\roman{enumi}}
\begin{Authlist}
\item \Adef{0}Deceased
\item \Adef{Dipartimento di Fisica dell'Universita, Udine, Italy}Also at: Dipartimento di Fisica dell'Universita, Udine, Italy
\item \Adef{M.V.Lomonosov Moscow State University, D.V.Skobeltsyn Institute of Nuclear Physics, Moscow, Russia}Also at: M.V.Lomonosov Moscow State University, D.V.Skobeltsyn Institute of Nuclear Physics, Moscow, Russia
\item \Adef{Institute of Nuclear Sciences, Belgrade, Serbia}Also at: "Vin\v{c}a" Institute of Nuclear Sciences, Belgrade, Serbia
\end{Authlist}
\section*{Collaboration Institutes}
\renewcommand\theenumi{\arabic{enumi}~}
\begin{Authlist}
\item \Idef{org1279}Benem\'{e}rita Universidad Aut\'{o}noma de Puebla, Puebla, Mexico
\item \Idef{org1220}Bogolyubov Institute for Theoretical Physics, Kiev, Ukraine
\item \Idef{org1262}Budker Institute for Nuclear Physics, Novosibirsk, Russia
\item \Idef{org1292}California Polytechnic State University, San Luis Obispo, California, United States
\item \Idef{org14939}Centre de Calcul de l'IN2P3, Villeurbanne, France
\item \Idef{org1197}Centro de Aplicaciones Tecnol\'{o}gicas y Desarrollo Nuclear (CEADEN), Havana, Cuba
\item \Idef{org1242}Centro de Investigaciones Energ\'{e}ticas Medioambientales y Tecnol\'{o}gicas (CIEMAT), Madrid, Spain
\item \Idef{org1244}Centro de Investigaci\'{o}n y de Estudios Avanzados (CINVESTAV), Mexico City and M\'{e}rida, Mexico
\item \Idef{org1335}Centro Fermi -- Centro Studi e Ricerche e Museo Storico della Fisica ``Enrico Fermi'', Rome, Italy
\item \Idef{org17347}Chicago State University, Chicago, United States
\item \Idef{org1118}China Institute of Atomic Energy, Beijing, China
\item \Idef{org1288}Commissariat \`{a} l'Energie Atomique, IRFU, Saclay, France
\item \Idef{org1294}Departamento de F\'{\i}sica de Part\'{\i}culas and IGFAE, Universidad de Santiago de Compostela, Santiago de Compostela, Spain
\item \Idef{org1106}Department of Physics Aligarh Muslim University, Aligarh, India
\item \Idef{org1121}Department of Physics and Technology, University of Bergen, Bergen, Norway
\item \Idef{org1162}Department of Physics, Ohio State University, Columbus, Ohio, United States
\item \Idef{org1300}Department of Physics, Sejong University, Seoul, South Korea
\item \Idef{org1268}Department of Physics, University of Oslo, Oslo, Norway
\item \Idef{org1132}Dipartimento di Fisica dell'Universit\`{a} and Sezione INFN, Bologna, Italy
\item \Idef{org1315}Dipartimento di Fisica dell'Universit\`{a} and Sezione INFN, Trieste, Italy
\item \Idef{org1145}Dipartimento di Fisica dell'Universit\`{a} and Sezione INFN, Cagliari, Italy
\item \Idef{org1270}Dipartimento di Fisica dell'Universit\`{a} and Sezione INFN, Padova, Italy
\item \Idef{org1285}Dipartimento di Fisica dell'Universit\`{a} `La Sapienza' and Sezione INFN, Rome, Italy
\item \Idef{org1154}Dipartimento di Fisica e Astronomia dell'Universit\`{a} and Sezione INFN, Catania, Italy
\item \Idef{org1290}Dipartimento di Fisica `E.R.~Caianiello' dell'Universit\`{a} and Gruppo Collegato INFN, Salerno, Italy
\item \Idef{org1312}Dipartimento di Fisica Sperimentale dell'Universit\`{a} and Sezione INFN, Turin, Italy
\item \Idef{org1103}Dipartimento di Scienze e Tecnologie Avanzate dell'Universit\`{a} del Piemonte Orientale and Gruppo Collegato INFN, Alessandria, Italy
\item \Idef{org1114}Dipartimento Interateneo di Fisica `M.~Merlin' and Sezione INFN, Bari, Italy
\item \Idef{org1237}Division of Experimental High Energy Physics, University of Lund, Lund, Sweden
\item \Idef{org1192}European Organization for Nuclear Research (CERN), Geneva, Switzerland
\item \Idef{org1227}Fachhochschule K\"{o}ln, K\"{o}ln, Germany
\item \Idef{org1122}Faculty of Engineering, Bergen University College, Bergen, Norway
\item \Idef{org1136}Faculty of Mathematics, Physics and Informatics, Comenius University, Bratislava, Slovakia
\item \Idef{org1274}Faculty of Nuclear Sciences and Physical Engineering, Czech Technical University in Prague, Prague, Czech Republic
\item \Idef{org1229}Faculty of Science, P.J.~\v{S}af\'{a}rik University, Ko\v{s}ice, Slovakia
\item \Idef{org1184}Frankfurt Institute for Advanced Studies, Johann Wolfgang Goethe-Universit\"{a}t Frankfurt, Frankfurt, Germany
\item \Idef{org1215}Gangneung-Wonju National University, Gangneung, South Korea
\item \Idef{org1212}Helsinki Institute of Physics (HIP) and University of Jyv\"{a}skyl\"{a}, Jyv\"{a}skyl\"{a}, Finland
\item \Idef{org1203}Hiroshima University, Hiroshima, Japan
\item \Idef{org1329}Hua-Zhong Normal University, Wuhan, China
\item \Idef{org1254}Indian Institute of Technology, Mumbai, India
\item \Idef{org1266}Institut de Physique Nucl\'{e}aire d'Orsay (IPNO), Universit\'{e} Paris-Sud, CNRS-IN2P3, Orsay, France
\item \Idef{org1277}Institute for High Energy Physics, Protvino, Russia
\item \Idef{org1249}Institute for Nuclear Research, Academy of Sciences, Moscow, Russia
\item \Idef{org1320}Nikhef, National Institute for Subatomic Physics and Institute for Subatomic Physics of Utrecht University, Utrecht, Netherlands
\item \Idef{org1250}Institute for Theoretical and Experimental Physics, Moscow, Russia
\item \Idef{org1230}Institute of Experimental Physics, Slovak Academy of Sciences, Ko\v{s}ice, Slovakia
\item \Idef{org1127}Institute of Physics, Bhubaneswar, India
\item \Idef{org1275}Institute of Physics, Academy of Sciences of the Czech Republic, Prague, Czech Republic
\item \Idef{org1139}Institute of Space Sciences (ISS), Bucharest, Romania
\item \Idef{org1185}Institut f\"{u}r Kernphysik, Johann Wolfgang Goethe-Universit\"{a}t Frankfurt, Frankfurt, Germany
\item \Idef{org1177}Institut f\"{u}r Kernphysik, Technische Universit\"{a}t Darmstadt, Darmstadt, Germany
\item \Idef{org1256}Institut f\"{u}r Kernphysik, Westf\"{a}lische Wilhelms-Universit\"{a}t M\"{u}nster, M\"{u}nster, Germany
\item \Idef{org1246}Instituto de Ciencias Nucleares, Universidad Nacional Aut\'{o}noma de M\'{e}xico, Mexico City, Mexico
\item \Idef{org1247}Instituto de F\'{\i}sica, Universidad Nacional Aut\'{o}noma de M\'{e}xico, Mexico City, Mexico
\item \Idef{org23333}Institut of Theoretical Physics, University of Wroclaw
\item \Idef{org1308}Institut Pluridisciplinaire Hubert Curien (IPHC), Universit\'{e} de Strasbourg, CNRS-IN2P3, Strasbourg, France
\item \Idef{org1182}Joint Institute for Nuclear Research (JINR), Dubna, Russia
\item \Idef{org1143}KFKI Research Institute for Particle and Nuclear Physics, Hungarian Academy of Sciences, Budapest, Hungary
\item \Idef{org1199}Kirchhoff-Institut f\"{u}r Physik, Ruprecht-Karls-Universit\"{a}t Heidelberg, Heidelberg, Germany
\item \Idef{org20954}Korea Institute of Science and Technology Information
\item \Idef{org1160}Laboratoire de Physique Corpusculaire (LPC), Clermont Universit\'{e}, Universit\'{e} Blaise Pascal, CNRS--IN2P3, Clermont-Ferrand, France
\item \Idef{org1194}Laboratoire de Physique Subatomique et de Cosmologie (LPSC), Universit\'{e} Joseph Fourier, CNRS-IN2P3, Institut Polytechnique de Grenoble, Grenoble, France
\item \Idef{org1187}Laboratori Nazionali di Frascati, INFN, Frascati, Italy
\item \Idef{org1232}Laboratori Nazionali di Legnaro, INFN, Legnaro, Italy
\item \Idef{org1125}Lawrence Berkeley National Laboratory, Berkeley, California, United States
\item \Idef{org1234}Lawrence Livermore National Laboratory, Livermore, California, United States
\item \Idef{org1251}Moscow Engineering Physics Institute, Moscow, Russia
\item \Idef{org1140}National Institute for Physics and Nuclear Engineering, Bucharest, Romania
\item \Idef{org1165}Niels Bohr Institute, University of Copenhagen, Copenhagen, Denmark
\item \Idef{org1109}Nikhef, National Institute for Subatomic Physics, Amsterdam, Netherlands
\item \Idef{org1283}Nuclear Physics Institute, Academy of Sciences of the Czech Republic, \v{R}e\v{z} u Prahy, Czech Republic
\item \Idef{org1264}Oak Ridge National Laboratory, Oak Ridge, Tennessee, United States
\item \Idef{org1189}Petersburg Nuclear Physics Institute, Gatchina, Russia
\item \Idef{org1170}Physics Department, Creighton University, Omaha, Nebraska, United States
\item \Idef{org1157}Physics Department, Panjab University, Chandigarh, India
\item \Idef{org1112}Physics Department, University of Athens, Athens, Greece
\item \Idef{org1152}Physics Department, University of Cape Town, iThemba LABS, Cape Town, South Africa
\item \Idef{org1209}Physics Department, University of Jammu, Jammu, India
\item \Idef{org1207}Physics Department, University of Rajasthan, Jaipur, India
\item \Idef{org1200}Physikalisches Institut, Ruprecht-Karls-Universit\"{a}t Heidelberg, Heidelberg, Germany
\item \Idef{org1325}Purdue University, West Lafayette, Indiana, United States
\item \Idef{org1281}Pusan National University, Pusan, South Korea
\item \Idef{org1176}Research Division and ExtreMe Matter Institute EMMI, GSI Helmholtzzentrum f\"ur Schwerionenforschung, Darmstadt, Germany
\item \Idef{org1334}Rudjer Bo\v{s}kovi\'{c} Institute, Zagreb, Croatia
\item \Idef{org1298}Russian Federal Nuclear Center (VNIIEF), Sarov, Russia
\item \Idef{org1252}Russian Research Centre Kurchatov Institute, Moscow, Russia
\item \Idef{org1224}Saha Institute of Nuclear Physics, Kolkata, India
\item \Idef{org1130}School of Physics and Astronomy, University of Birmingham, Birmingham, United Kingdom
\item \Idef{org1338}Secci\'{o}n F\'{\i}sica, Departamento de Ciencias, Pontificia Universidad Cat\'{o}lica del Per\'{u}, Lima, Peru
\item \Idef{org1146}Sezione INFN, Cagliari, Italy
\item \Idef{org1155}Sezione INFN, Catania, Italy
\item \Idef{org1286}Sezione INFN, Rome, Italy
\item \Idef{org1115}Sezione INFN, Bari, Italy
\item \Idef{org1271}Sezione INFN, Padova, Italy
\item \Idef{org1313}Sezione INFN, Turin, Italy
\item \Idef{org1133}Sezione INFN, Bologna, Italy
\item \Idef{org1316}Sezione INFN, Trieste, Italy
\item \Idef{org1322}Soltan Institute for Nuclear Studies, Warsaw, Poland
\item \Idef{org1258}SUBATECH, Ecole des Mines de Nantes, Universit\'{e} de Nantes, CNRS-IN2P3, Nantes, France
\item \Idef{org1304}Technical University of Split FESB, Split, Croatia
\item \Idef{org1168}The Henryk Niewodniczanski Institute of Nuclear Physics, Polish Academy of Sciences, Cracow, Poland
\item \Idef{org17361}The University of Texas at Austin, Physics Department, Austin, TX, United States
\item \Idef{org1173}Universidad Aut\'{o}noma de Sinaloa, Culiac\'{a}n, Mexico
\item \Idef{org1296}Universidade de S\~{a}o Paulo (USP), S\~{a}o Paulo, Brazil
\item \Idef{org1149}Universidade Estadual de Campinas (UNICAMP), Campinas, Brazil
\item \Idef{org1239}Universit\'{e} de Lyon, Universit\'{e} Lyon 1, CNRS/IN2P3, IPN-Lyon, Villeurbanne, France
\item \Idef{org1205}University of Houston, Houston, Texas, United States
\item \Idef{org20371}University of Technology and Austrian Academy of Sciences, Vienna, Austria
\item \Idef{org1222}University of Tennessee, Knoxville, Tennessee, United States
\item \Idef{org1310}University of Tokyo, Tokyo, Japan
\item \Idef{org1318}University of Tsukuba, Tsukuba, Japan
\item \Idef{org21360}Eberhard Karls Universit\"{a}t T\"{u}bingen, T\"{u}bingen, Germany
\item \Idef{org1225}Variable Energy Cyclotron Centre, Kolkata, India
\item \Idef{org1306}V.~Fock Institute for Physics, St. Petersburg State University, St. Petersburg, Russia
\item \Idef{org1323}Warsaw University of Technology, Warsaw, Poland
\item \Idef{org1179}Wayne State University, Detroit, Michigan, United States
\item \Idef{org1260}Yale University, New Haven, Connecticut, United States
\item \Idef{org1332}Yerevan Physics Institute, Yerevan, Armenia
\item \Idef{org15649}Yildiz Technical University, Istanbul, Turkey
\item \Idef{org1301}Yonsei University, Seoul, South Korea
\item \Idef{org1327}Zentrum f\"{u}r Technologietransfer und Telekommunikation (ZTT), Fachhochschule Worms, Worms, Germany
\end{Authlist}
\endgroup

\else
\ifbibtex
\bibliographystyle{unsrtnat}
\bibliography{biblio}{}
\else

\fi
\fi
\else
\iffull

\input{refpaper.tex}
\else
\ifbibtex
\bibliographystyle{model1-num-names}
\bibliography{biblio}{}

\begin{thebibliography}{45}
\providecommand{\natexlab}[1]{#1}
\providecommand{\url}[1]{\texttt{#1}}
\expandafter\ifx\csname urlstyle\endcsname\relax
  \providecommand{\doi}[1]{doi: #1}\else
  \providecommand{\doi}{doi: \begingroup \urlstyle{rm}\Url}\fi

\bibitem[Arsene et~al.(2005)]{Arsene:2004fa}
I.\ Arsene et~al.
\newblock {Quark Gluon Plasma an Color Glass Condensate at RHIC? The
  perspective from the BRAHMS experiment}.
\newblock \emph{Nucl.~Phys.}, A757:\penalty0 1--27, 2005.
\newblock \doi{10.1016/j.nuclphysa.2005.02.130}.

\bibitem[Back et~al.(2005)]{Back:2004je}
B.~B.\ Back et~al.
\newblock {The PHOBOS perspective on discoveries at RHIC}.
\newblock \emph{Nucl.~Phys.}, A757:\penalty0 28--101, 2005.
\newblock \doi{10.1016/j.nuclphysa.2005.03.084}.

\bibitem[Adams et~al.(2005{\natexlab{a}})]{Adams:2005dq}
J.\ Adams et~al.
\newblock {Experimental and theoretical challenges in the search for the quark
  gluon plasma: The STAR collaboration's critical assessment of the evidence
  from RHIC collisions}.
\newblock \emph{Nucl.~Phys.}, A757:\penalty0 102--183, 2005{\natexlab{a}}.
\newblock \doi{10.1016/j.nuclphysa.2005.03.085}.

\bibitem[Adcox et~al.(2005)]{Adcox:2004mh}
K.\ Adcox et~al.
\newblock {Formation of dense partonic matter in relativistic nucleus nucleus
  collisions at RHIC: Experimental evaluation by the PHENIX collaboration}.
\newblock \emph{Nucl.~Phys.}, A757:\penalty0 184--283, 2005.
\newblock \doi{10.1016/j.nuclphysa.2005.03.086}.

\bibitem[Aamodt et~al.(2010)]{Aamodt:2010pa}
K.\ Aamodt et~al.
\newblock {Elliptic flow of charged particles in Pb+Pb collisions at 2.76 TeV}.
\newblock \emph{Phys.~Rev.~Lett.}, 105:\penalty0 252302, 2010.
\newblock \doi{10.1103/PhysRevLett.105.252302}.

\bibitem[Collaboration(2011)]{Collaboration:2011yk}
ATLAS Collaboration.
\newblock {Measurement of the pseudorapidity and transverse momentum dependence
  of the elliptic flow of charged particles in lead-lead collisions at
  $\sqrt{s_{_{NN}}}$ = 2.76 TeV with the ATLAS detector}.
\newblock \emph{\arxiv{1108.6018} (hep-ex)}, 2011.
\newblock * Temporary entry *.

\bibitem[Adler et~al.(2003)]{Adler:2002tq}
C.\ Adler et~al.
\newblock {Disappearance of back-to-back high p(T) hadron correlations in
  central Au+Au collisions at $\snn = 200$ GeV}.
\newblock \emph{Phys.~Rev.~Lett.}, 90:\penalty0 082302, 2003.
\newblock \doi{10.1103/PhysRevLett.90.082302}.

\bibitem[Adams et~al.(2005{\natexlab{b}})]{Adams:2005ph}
J.\ Adams et~al.
\newblock {Distributions of charged hadrons associated with high transverse
  momentum particles in p p and Au + Au collisions at $\snn = 200$ GeV}.
\newblock \emph{Phys.~Rev.~Lett.}, 95:\penalty0 152301, 2005{\natexlab{b}}.
\newblock \doi{10.1103/PhysRevLett.95.152301}.

\bibitem[Adare et~al.(2007)]{Adare:2006nr}
A.\ Adare et~al.
\newblock {System size and energy dependence of jet-induced hadron pair
  correlation shapes in Cu + Cu and Au + Au collisions at $\snn = $ 200 GeV and
  62.4 GeV}.
\newblock \emph{Phys.~Rev.~Lett.}, 98:\penalty0 232302, 2007.
\newblock \doi{10.1103/PhysRevLett.98.232302}.

\bibitem[Adams et~al.(2006)]{Adams:2006yt}
J.\ Adams et~al.
\newblock {Direct observation of dijets in central Au + Au collisions at $\snn
  = 200$ GeV}.
\newblock \emph{Phys.~Rev.~Lett.}, 97:\penalty0 162301, 2006.
\newblock \doi{10.1103/PhysRevLett.97.162301}.

\bibitem[Alver et~al.(2010{\natexlab{a}})]{Alver:2008gk}
B.\ Alver et~al.
\newblock {System size dependence of cluster properties from two- particle
  angular correlations in Cu+Cu and Au+Au collisions at $\sqrt{s_{_{NN}}}$ =
  200 GeV}.
\newblock \emph{Phys.~Rev.}, C81:\penalty0 024904, 2010{\natexlab{a}}.
\newblock \doi{10.1103/PhysRevC.81.024904}.

\bibitem[Adare et~al.(2008)]{Adare:2008cqb}
A.\ Adare et~al.
\newblock {Dihadron azimuthal correlations in Au+Au collisions at $\snn = 200$
  GeV}.
\newblock \emph{Phys.Rev.}, C78:\penalty0 014901, 2008.
\newblock \doi{10.1103/PhysRevC.78.014901}.

\bibitem[Abelev et~al.(2009)]{Abelev:2009qa}
B.~I.\ Abelev et~al.
\newblock {Long range rapidity correlations and jet production in high energy
  nuclear collisions}.
\newblock \emph{Phys.~Rev.}, C80:\penalty0 064912, 2009.
\newblock \doi{10.1103/PhysRevC.80.064912}.

\bibitem[Alver et~al.(2010{\natexlab{b}})]{Alver:2009id}
B.\ Alver et~al.
\newblock {High transverse momentum triggered correlations over a large
  pseudorapidity acceptance in Au+Au collisions at $\snn=200$ GeV}.
\newblock \emph{Phys.~Rev.~Lett.}, 104:\penalty0 062301, 2010{\natexlab{b}}.
\newblock \doi{10.1103/PhysRevLett.104.062301}.

\bibitem[Adare et~al.(2010)]{Adare:2010ry}
A.\ Adare et~al.
\newblock {Trends in Yield and Azimuthal Shape Modification in Dihadron
  Correlations in Relativistic Heavy Ion Collisions}.
\newblock \emph{Phys.~Rev.~Lett.}, 104:\penalty0 252301, 2010.
\newblock \doi{10.1103/PhysRevLett.104.252301}.

\bibitem[Khachatryan et~al.(2010)]{Khachatryan:2010gv}
V.\ Khachatryan et~al.
\newblock {Observation of Long-Range Near-Side Angular Correlations in
  Proton-Proton Collisions at the LHC}.
\newblock \emph{JHEP}, 09:\penalty0 091, 2010.
\newblock \doi{10.1007/JHEP09(2010)091}.

\bibitem[Aad et~al.(2011)]{ATLAS-CONF-2011-074}
C.\ Aad et~al.
\newblock Measurement of elliptic flow and higher-order flow coefficients with
  the atlas detector in $\snn=2.76$ tev pb+pb collisions.
\newblock Technical Report ATLAS-CONF-2011-074, CERN, Geneva, May 2011.

\bibitem[Chatrchyan et~al.(2011)]{Chatrchyan:2011ek}
S.\ Chatrchyan et~al.
\newblock {Long-range and short-range dihadron angular correlations in central
  PbPb collisions at a nucleon-nucleon center of mass energy of 2.76 TeV}.
\newblock \emph{\arxiv{1105.2438} (nucl-ex)}, 2011.

\bibitem[Voloshin(2006)]{Voloshin:2006ei}
S~Voloshin.
\newblock {Transverse radial expansion in nuclear collisions and two particle
  correlations}.
\newblock \emph{Physics Letters B}, 632\penalty0 (4):\penalty0 490--494,
  January 2006.

\bibitem[Armesto et~al.(2004)Armesto, Salgado, and Wiedemann]{Armesto:2004pt}
N.\ Armesto, C.~A.\ Salgado, and U.~A. Wiedemann.
\newblock {Measuring the collective flow with jets}.
\newblock \emph{Phys.~Rev.~Lett.}, 93:\penalty0 242301, 2004.
\newblock \doi{10.1103/PhysRevLett.93.242301}.

\bibitem[Chiu and Hwa(2005)]{Chiu:2005ad}
C.~B.\ Chiu and R.~C. Hwa.
\newblock {Pedestal and peak structure in jet correlation}.
\newblock \emph{Phys.~Rev.}, C72:\penalty0 034903, 2005.
\newblock \doi{10.1103/PhysRevC.72.034903}.

\bibitem[Romatschke(2007)]{Romatschke:2006bb}
P.~Romatschke.
\newblock {Momentum broadening in an anisotropic plasma}.
\newblock \emph{Phys.~Rev.}, C75:\penalty0 014901, 2007.
\newblock \doi{10.1103/PhysRevC.75.014901}.

\bibitem[Majumder et~al.(2007)Majumder, Muller, and Bass]{Majumder:2006wi}
A.\ Majumder, B.\ Muller, and S.~A. Bass.
\newblock {Longitudinal Broadening of Quenched Jets in Turbulent Color Fields}.
\newblock \emph{Phys.~Rev.~Lett.}, 99:\penalty0 042301, 2007.
\newblock \doi{10.1103/PhysRevLett.99.042301}.

\bibitem[Shuryak(2007)]{Shuryak:2007fu}
E.~V. Shuryak.
\newblock {On the Origin of the 'Ridge' phenomenon induced by Jets in Heavy Ion
  Collisions}.
\newblock \emph{Phys.~Rev.}, C76:\penalty0 047901, 2007.
\newblock \doi{10.1103/PhysRevC.76.047901}.

\bibitem[Wong(2008)]{Wong:2008yh}
C.-Y. Wong.
\newblock {The Momentum Kick Model Description of the Near-Side Ridge and Jet
  Quenching}.
\newblock \emph{Phys.~Rev.}, C78:\penalty0 064905, 2008.
\newblock \doi{10.1103/PhysRevC.78.064905}.

\bibitem[Dumitru et~al.(2008)Dumitru, Gelis, McLerran, and
  Venugopalan]{Dumitru:2008wn}
A.\ Dumitru, F.\ Gelis, L.\ McLerran, and R.~Venugopalan.
\newblock {Glasma flux tubes and the near side ridge phenomenon at RHIC}.
\newblock \emph{Nucl.~Phys.}, A810:\penalty0 91--108, 2008.
\newblock \doi{10.1016/j.nuclphysa.2008.06.012}.

\bibitem[Gavin et~al.(2009)Gavin, McLerran, and Moschelli]{Gavin:2008ev}
S.\ Gavin, L.\ McLerran, and G.~Moschelli.
\newblock {Long Range Correlations and the Soft Ridge in Relativistic Nuclear
  Collisions}.
\newblock \emph{Phys.~Rev.}, C79:\penalty0 051902, 2009.
\newblock \doi{10.1103/PhysRevC.79.051902}.

\bibitem[Dusling et~al.(2009)Dusling, Fernandez-Fraile, and
  Venugopalan]{Dusling:2009ar}
K.\ Dusling, D.\ Fernandez-Fraile, and R.~Venugopalan.
\newblock {Three-particle correlation from glasma flux tubes}.
\newblock \emph{Nucl.~Phys.}, A828:\penalty0 161--177, 2009.
\newblock \doi{10.1016/j.nuclphysa.2009.06.017}.

\bibitem[Hama et~al.(2009)Hama, Andrade, Grassi, and Qian]{Hama:2009vu}
Y.\ Hama, R.~P.~G.\ Andrade, F.\ Grassi, and W.-L. Qian.
\newblock {Trying to understand the ridge effect in hydrodynamic model}.
\newblock \emph{Nonlin.~Phenom.~Complex Syst.}, 12:\penalty0 466--470, 2009.

\bibitem[Morsch(2006)]{Morsch:2006pf}
A.~Morsch.
\newblock {On the mean parton transverse momentum versus associated hadron p(T)
  in di-hadron correlations at RHIC and LHC}.
\newblock \emph{\arxiv{hep-ph/0606098}}, 2006.

\bibitem[Aamodt et~al.(2011{\natexlab{a}})]{Aamodt:2011vk}
K.\ Aamodt et~al.
\newblock {Higher harmonic anisotropic flow measurements of charged particles
  in Pb+Pb collisions at 2.76 TeV}.
\newblock \emph{Phys.~Rev.~Lett.}, 107\penalty0 (3):\penalty0 032301, Jul
  2011{\natexlab{a}}.
\newblock \doi{10.1103/PhysRevLett.107.032301}.

\bibitem[Manly et~al.(2006)]{Manly:2005zy}
S.\ Manly et~al.
\newblock {System size, energy and pseudorapidity dependence of directed and
  elliptic flow at RHIC}.
\newblock \emph{Nucl.~Phys.}, A774:\penalty0 523--526, 2006.
\newblock \doi{10.1016/j.nuclphysa.2006.06.079}.

\bibitem[Alver et~al.(2007)]{Alver:2006wh}
B.\ Alver et~al.
\newblock {System size, energy, pseudorapidity, and centrality dependence of
  elliptic flow}.
\newblock \emph{Phys.~Rev.~Lett.}, 98:\penalty0 242302, 2007.
\newblock \doi{10.1103/PhysRevLett.98.242302}.

\bibitem[Mishra et~al.(2008)Mishra, Mohapatra, Saumia, and
  Srivastava]{Mishra:2007tw}
A.~P.\ Mishra, R.~K.\ Mohapatra, P.~S. Saumia, and A.~M. Srivastava.
\newblock {Super-horizon fluctuations and acoustic oscillations in relativistic
  heavy-ion collisions}.
\newblock \emph{Phys.~Rev.}, C77:\penalty0 064902, 2008.
\newblock \doi{10.1103/PhysRevC.77.064902}.

\bibitem[Mishra et~al.(2010)Mishra, Mohapatra, Saumia, and
  Srivastava]{Mishra:2008dm}
A.~P.\ Mishra, R.~K.\ Mohapatra, P.~S. Saumia, and A.~M. Srivastava.
\newblock {Using CMBR analysis tools for flow anisotropies in relativistic
  heavy-ion collisions}.
\newblock \emph{Phys.~Rev.}, C81:\penalty0 034903, 2010.
\newblock \doi{10.1103/PhysRevC.81.034903}.

\bibitem[Takahashi et~al.(2009)]{Takahashi:2009na}
J.\ Takahashi et~al.
\newblock {Topology studies of hydrodynamics using two particle correlation
  analysis}.
\newblock \emph{Phys.~Rev.~Lett.}, 103:\penalty0 242301, 2009.
\newblock \doi{10.1103/PhysRevLett.103.242301}.

\bibitem[Sorensen(2010)]{Sorensen:2010sqm}
Paul Sorensen.
\newblock Implications of space-momentum correlations and geometric
  fluctuations in heavy-ion collisions.
\newblock \emph{Journal of Physics G: Nuclear and Particle Physics},
  37\penalty0 (9):\penalty0 094011, 2010.
\newblock URL \url{http://stacks.iop.org/0954-3899/37/i=9/a=094011}.

\bibitem[Alver and Roland(2010)]{Alver:2010gr}
B.\ Alver and G.~Roland.
\newblock {Collision geometry fluctuations and triangular flow in heavy-ion
  collisions}.
\newblock \emph{Phys.~Rev.}, C81:\penalty0 054905, 2010.
\newblock \doi{10.1103/PhysRevC.81.054905}.

\bibitem[Teaney and Yan(2010)]{Teaney:2010vd}
D.\ Teaney and L.~Yan.
\newblock {Triangularity and Dipole Asymmetry in Heavy Ion Collisions}.
\newblock \emph{\arxiv{1010.1876} (nucl-th)}, 2010.

\bibitem[Luzum(2011{\natexlab{a}})]{Luzum:2010sp}
M.~Luzum.
\newblock {Collective flow and long-range correlations in relativistic heavy
  ion collisions}.
\newblock \emph{Phys.~Lett.}, B696:\penalty0 499--504, 2011{\natexlab{a}}.
\newblock \doi{10.1016/j.physletb.2011.01.013}.

\bibitem[Adare et~al.(2011)]{Adare:2011tg}
A.\ Adare et~al.
\newblock {Measurements of Higher-Order Flow Harmonics in Au+Au Collisions at
  $\snn= 200$ GeV}.
\newblock \emph{\arxiv{1105.3928} (nucl-ex)}, 2011.

\bibitem[Aamodt et~al.(2011{\natexlab{b}})]{Aamodt:2010cz}
K.\ Aamodt et~al.
\newblock {Centrality dependence of the charged-particle multiplicity density
  at mid-rapidity in Pb-Pb collisions at $\snn = 2.76$ TeV}.
\newblock \emph{Phys.~Rev.~Lett.}, 106:\penalty0 032301, 2011{\natexlab{b}}.

\bibitem[Gardim et~al.(2011)Gardim, Grassi, Hama, Luzum, and
  Ollitrault]{Gardim:2011qn}
F.~G.\ Gardim, F.\ Grassi, Y.~Hama, M.\ Luzum, and J.-Y. Ollitrault.
\newblock {Directed flow at mid-rapidity in event-by-event hydrodynamics}.
\newblock \emph{Phys.~Rev.}, C83:\penalty0 064901, 2011.

\bibitem[Luzum(2011{\natexlab{b}})]{luzum_privcomm}
M.~Luzum.
\newblock \emph{private communication}, 2011{\natexlab{b}}.

\bibitem[Casalderrey-Solana et~al.(2006)Casalderrey-Solana, Shuryak, and
  Teaney]{CasalderreySolana:2004qm}
J.\ Casalderrey-Solana, E.\~V.\ Shuryak, and D.~Teaney.
\newblock {Conical flow induced by quenched QCD jets}.
\newblock \emph{Nucl.~Phys.}, A774:\penalty0 577, 2006.
\newblock \doi{10.1016/j.nuclphysa.2006.06.091}.
\newblock [Nucl.Phys.A774:577-580,2006].

\end{thebibliography}
\else
\input{refpaper.tex}
\fi
\fi
\fi
\end{document}